\documentclass{article}

\usepackage[a4paper, margin=1.0in]{geometry}
\usepackage[utf8]{inputenc}
\usepackage[dvipsnames,table,xcdraw]{xcolor}
\usepackage[affil-it]{authblk}
\usepackage[misc]{ifsym}
\usepackage[mathscr]{euscript}
\usepackage{algorithm}
\usepackage{algpseudocode}

\usepackage{appendix,
verbatim,
listings,
braket,
enumitem, 
kantlipsum,
graphicx,
multirow,
multicol,
mathtools,
csquotes,
amsmath,
mathtools,
url,
subfiles,
qcircuit,
amsthm,
amssymb,
fdsymbol,
rotating,
caption,
hyperref,
fontawesome,
xcolor,
sectsty,
arydshln,
ulem,
subcaption,
caption,
array}

\def\titlefont{\color{RoyalPurple}}
\sectionfont{\color{RoyalPurple}}
\subsectionfont{\color{RoyalPurple}}
\subsubsectionfont{\color{RoyalPurple}}

\NewDocumentCommand{\rot}{O{45} O{1em} m}{\makebox[#2][l]{\rotatebox{#1}{#3}}}

\let\OLDthebibliography\thebibliography
\renewcommand\thebibliography[1]{
  \OLDthebibliography{#1}
  \setlength{\parskip}{0pt}
}

\title{\titlefont \textbf{\large EO-GRAPE and EO-DRLPE: Open and Closed Loop Approaches\\ for Energy Efficient Quantum Optimal Control}}

\author[1,2]{Sebastiaan Fauquenot}  
\author[1,2]{Aritra Sarkar}
\author[1,2]{Sebastian Feld}

\affil[1]{Quantum Machine Learning research group,
Quantum Computing division, QuTech, The Netherlands}
\affil[2]{Department of Quantum \& Computer Engineering, Delft University of Technology, The Netherlands}

\affil[ \Letter ]{s.feld@tudelft.nl}

\date{}

\begin{document}

\maketitle
\begin{abstract} 
This research investigates the possibility of using quantum optimal control techniques to co-optimize the energetic cost and the process fidelity of a quantum unitary gate.
The energetic cost is theoretically defined, and thereby, the gradient of the energetic cost for pulse engineering is derived.
We empirically demonstrate the Pareto optimality in the trade-off between process fidelity and energetic cost. 
Thereafter, two novel numerical quantum optimal control approaches are proposed: (i) energy-optimized gradient ascent pulse engineering~(EO-GRAPE) as an open-loop gradient-based method, and (ii) energy-optimized deep reinforcement learning for pulse engineering~(EO-DRLPE) as a closed-loop method. 
The performance of both methods is probed in the presence of increasing noise.  
We find that the EO-GRAPE method performs better than the EO-DRLPE methods with and without a warm start for most experimental settings.
Additionally, for one qubit unitary gate, we illustrate the correlation between the Bloch sphere path length and the energetic cost.
\end{abstract}

\section{Introduction} \label{sec:introduction}

The field of quantum computing~(QC) is undergoing rapid development in both theoretical and practical aspects from both academic and industrial stakeholders.
At its foundation, QC involves orchestrating quantum mechanical properties for information processing~\cite{benioff1980computer,deutsch1985quantum} via quantum algorithms, thereby promising a significant reduction in computational resources~\cite{bernstein1993quantum} for specific applications over their classical counterparts.
Exemplary use cases include simulating quantum mechanical systems~\cite{feynman1982simulating} towards novel discovery in material sciences and breaking cryptographic protocols~\cite{shor2022early} ubiquitous for secure transactions over the internet.
The required orchestration of quantum information towards QC is physically implemented by engineering quantum processing units~(QPU), which are currently prototyped using a myriad of technologies like superconducting circuits, trapped ions, photonics, electron spin, etc.
A significant challenge in the QC field~\cite{leymann2020bitter,ezratty2023we,waintal2024quantum} is to manufacture better quality and scalable QPU against the fragility of quantum information from environmental noises and operational imperfections.

\begin{figure}[htb]
    \centering 
    \includegraphics[clip, trim=6.1cm 1.0cm 1.0cm 36.2cm, width = 1.0\linewidth]{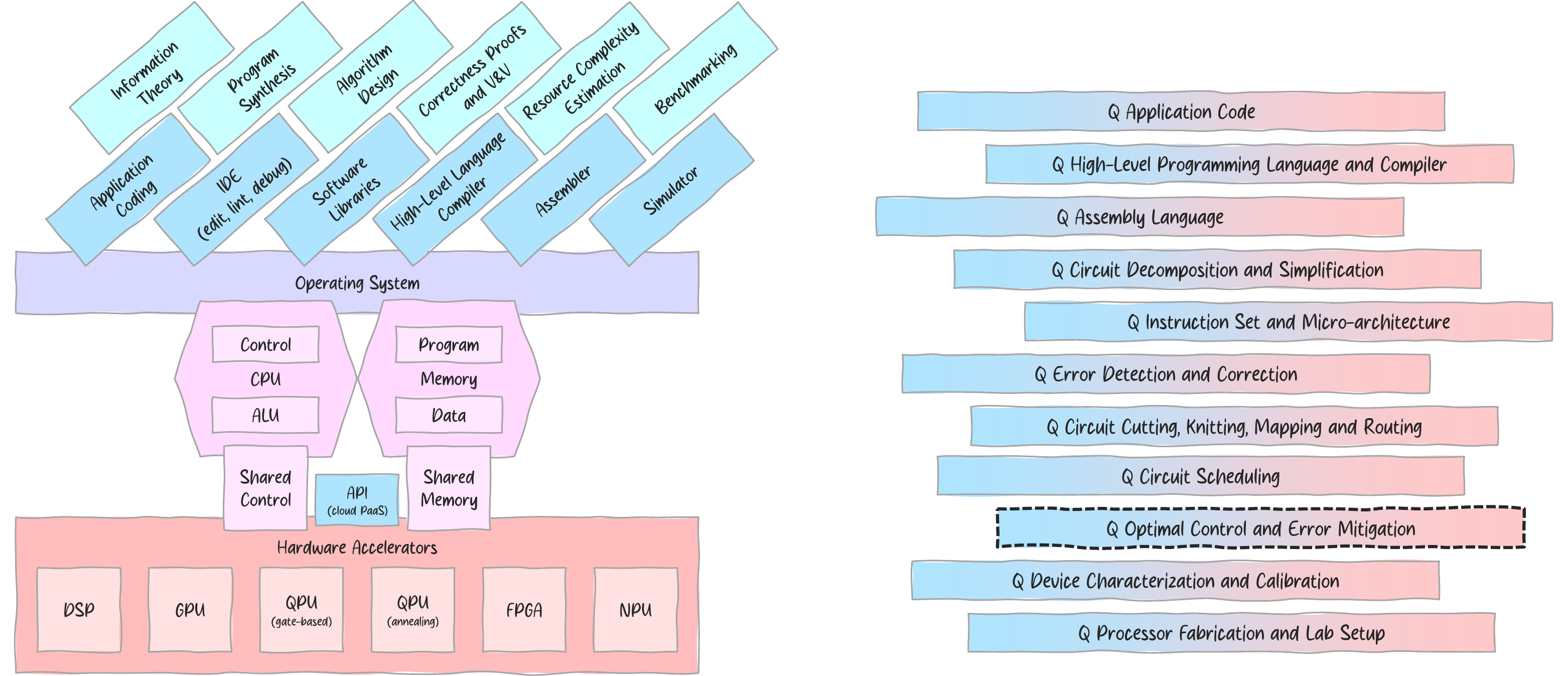}
    \caption{Overview of the system design of a quantum accelerator with classical control and various software modules required for research and development is shown on the left. The different abstraction layers for full-stack quantum computing are shown on the right. This research pertains to the optimal control layer highlighted with a dotted outline.}
    \label{fig:qcstack}
\end{figure}

Akin to classical computers, the interface between the QC applications and the processor is organized into translation layers, called the quantum computation stack~\cite{bertels2020quantum,bertels2021quantum}, as shown in figure \ref{fig:qcstack}.
From top to bottom, first, the application is formulated as a quantum algorithm and expressed in a quantum programming language.
Then, a quantum compiler decomposes and optimizes the high-level code into native operations supported by the target QPU.
Thereafter, the quantum microarchitecture schedules and issues the low-level instructions in real-time.
These instructions (like initialization, unitary gates, and measurements) further need to be translated to corresponding analog pulses that optimally control the accessible degrees of freedom of the quantum system.
These electromagnetic analog signals perform the necessary transformation for synthesizing quantum unitary gates on specific addressable qubits while mitigating the undesirable effects of noise.
Eventually, these hardware-aware signals implement the desired logical operation dictated by the hardware-agnostic quantum algorithm on the target QPU.

The QC stack, as discussed above, provides abstraction layers to organize the research and development.
Thus, operations, including a computationally universal set of gates, initialization, and measurements, are provided as primitives for the microarchitecture layer.
In the ideal scenario, the pulse-level implementation details of these operations are predetermined in the control layer during the characterization and calibration of the QPU prior to deployment.
However, in the current noisy intermediate-scale quantum~(NISQ) computing era, this abstraction~\cite{shi2020resource} is preventing further optimization of the QPU performance.
Pulse-level control, when exposed to higher stack layers, allows advanced operating procedures that can mitigate the effect of noise dynamics and fine-tune a more extensive set of unitary gates with respect to other dependencies (e.g., cross-talk).
While there are many advantages to pulse control, such as a high degree of flexibility and higher speed of gate execution, there are also certain downsides to using pulse control, such as calibration requirements and overhead and scalability bottlenecks.
Nevertheless, advancing research in this direction is not only imperative for current QC roadmaps to extract maximal performance but also allows a principled theoretical understanding of the limits of quantum information processing using classical control algorithms and electronics. 

Current pulse control approaches focus primarily on optimizing the fidelity score.
Motivated by the theoretical possibility and operational need, this research concerns a novel approach for pulse-level unitary gate synthesis, that co-optimizes the fidelity along with the pulse energy.
Within the allied field of quantum thermodynamics, there is a growing interest in the possibility of achieving quantum advantage through the perspective of energy efficiency~\cite{manifesto} instead of computational resources like runtime and memory needs.
This is imperative for at least two specific reasons.
Firstly, one of the earliest motivations of quantum computation stems from reversible computation~\cite{frank2018back} with minimal energetic cost instead of a computational complexity advantage.
Secondly, quantum processors within a quantum accelerator stack are benchmarked in performance against state-of-the-art classical high-performance computing~(HPC) systems.
In such HPC systems, optimizing and benchmarking the energetic cost~\cite{benhari2024green} is motivated by both operational costs and environmental sustainability directives.   
While this direction is gaining traction, specific research in the energetic cost of a quantum computational process is both sparse and theoretical at present. 
For example, \cite{deffner2021energeticcost} proves an inequality bounding the change of Shannon information encoded in the logical quantum states by quantifying the energetic cost of Hamiltonian gate operations. Subsequently, \cite{aifer2022quantum} showed that optimal control problems can be solved within the powerful framework on quantum speed limits and derive state-independent lower bounds on energetic cost. 
Recent work in quantum optimal control~(QOC) theory has primarily focused on developing control to carry out quantum processes with the highest fidelity possible. 
These processes include quantum processes such as state initialization, quantum measurements, and implementing quantum unitary gates. 
However, in the context of the growing impetus in achieving quantum advantage through energy efficiency~\cite{auffeves2022quantum}, it seems crucial to investigate the energy efficiency of quantum operations, in particular, unitary quantum gates within a pragmatic quantum compiler framework.
To this effect, we address three main research questions in this article: 
\begin{enumerate}[nolistsep,noitemsep,leftmargin=1.2cm]
    \item[$\mathcal{RQ}_1$:] How can we estimate the energetic cost of synthesizing a quantum unitary gate?
    \item[$\mathcal{RQ}_2$:] What is the relation between the fidelity of unitary synthesis and the energetic cost associated with the pulse?
    \item[$\mathcal{RQ}_3$:] How can fidelity and the energetic cost be co-optimized within existing quantum optimal control strategies?
\end{enumerate}

The core contributions of this research are summarized below:  
\begin{enumerate}[nolistsep,noitemsep]
\item A theoretical formulation of the energetic cost of implementing a quantum unitary gate using discrete control pulses and the gradient of the energetic cost with respect to the control parameters that are required to optimize the cost.
\item Development of a modified version of the gradient ascent pulse engineering~(GRAPE) algorithm to co-optimize a quantum unitary gate's fidelity and energetic cost. The proposed novel energy-optimized algorithm is called EO-GRAPE.
\item Identification and analysis of the trade-off between the fidelity and energetic cost of implementing a quantum unitary gate.
\item Development of a deep reinforcement learning~(DRL) agent able to learn and generate energy-optimized control pulses for a universal set of quantum unitary gates. The proposed novel energy-optimized method is called EO-DRLPE.
\item Benchmarks to evaluate the performance of the EO-GRAPE algorithm and EO-DRLPE method with increasing noise levels. 
\item Evaluation of the optimality of the geodesic path cost of the two methods for synthesizing 1-qubit unitary gates. 
\end{enumerate}

The rest of the article is organized as follows.
Section~\ref{sec:background} provides background information on pulse engineering for unitary synthesis. Thereafter, a non-exhaustive survey and classification of quantum optimal control techniques is provided. 
The energetic cost of a quantum operation is introduced.
In section~\ref{sec:eogrape}, the GRAPE algorithm is introduced. Thereafter, we present our novel EO-GRAPE algorithm that co-optimizes the energy and process fidelity based on the derived gradient. The trade-off between these factors is explored.
Section~\ref{sec:eo-drlpe} introduces the alternative closed-loop approach, EO-DRLPE and its consecutive performance analysis.
In section~\ref{sec:geodesic}, we provide an analysis of the correlation between the energetic cost and Bloch sphere path length for single qubit unitary synthesis.
Section~\ref{sec:conclusion} concludes the article and provides suggestions for future research directions.

\section{Background and related work} \label{sec:background}

In this section, the formulation of pulse-level unitary gate synthesis is presented.
Thereafter, a brief overview of existing QOC techniques is discussed to highlight the rationale behind the two methods that are proposed for energetic cost optimization.

\subsection{Pulse-level unitary gate synthesis}

Quantum systems evolve over time according to the time-dependent Schrödinger equation~(TDSE)~\cite{schrodinger1926undulatory},
\begin{equation}
    i \hbar \frac{d }{dt} | \psi (t) \rangle = \hat{H} | \psi (t) \rangle
    \label{eq:TDSE}
\end{equation}
where, $\hat{H}$ represents the Hamiltonian of the quantum system. 
A general solution for this equation is given by,
\begin{equation}
    | \psi (t) \rangle = e^{-i \hat{H}t/\hbar} | \psi (t_0) \rangle \equiv  U(t)| \psi (t_0) \rangle
\end{equation}
where,
\begin{equation}
    U(t) = e^{-i \hat{H}t/\hbar}
\end{equation}
$\hbar = \frac{h}{2\pi}$ is the reduced Planck constant and is typically set to $1$ in natural units.

Quantum logic gates are implemented by carefully tuning the Hamiltonian of the quantum system over time. 
Thus, if we want to perform a specific unitary (such as a Hadamard or CNOT), we have to apply a specific aggregated Hamiltonian to the system. 
This aggregated Hamiltonian is effectuated over a specific duration by carefully tuning the accessible/controllable Hamiltonians of the system over that duration.

We oftentimes make a distinction between the so-called drift Hamiltonian $\hat{H}_D$ and control Hamiltonian $\hat{H}_C$. 
The drift Hamiltonian pertains to the actual qubit(s), usually consisting of an individual term for the eigen-energy and eigenstates of the qubits, as well as a coupling term between different qubits. 
The control Hamiltonian describes the external control fields that can be applied to the qubit(s). 

Let us consider a simple case of a single qubit without any interaction terms. 
The qubit is described by the following Hamiltonian,
\begin{equation}
    \hat{H}_D = \hbar \omega_{0} \hat{\sigma}_{z}
\end{equation}
This implies the qubit will precess about the $\hat{z}$-axis with frequency $\omega_{0}$. 

If the control field is applied on the $\hat{x} \hat{y}$-plane, and rotates around the $\hat{z}$-axis with frequency $\omega_{rf}$, we can define the control Hamiltonian by,
\begin{equation}
    \hat{H}_C = \hbar \omega_1 \left ( \cos{\omega_{rf} t} \hat{\sigma}_x +\sin{\omega_{rf} t}  \hat{\sigma}_y \right )
\end{equation}
and we can thus rewrite our total Hamiltonian as,
\begin{equation}
    \hat{H} = \hat{H}_D + \hat{H}_C =\hbar \omega_0 \hat{\sigma}_z + \hbar \omega_1(\cos{\omega_{rf} t} \hat{\sigma}_x+\sin{\omega_{rf} t}  \hat{\sigma}_y)
    \label{eq:rotham}
\end{equation}

The control fields or the time-dependent function of the control Hamiltonian operators are referred to as the control pulses that one can apply to the quantum system. 
A control pulse field $k$ typically has three associated parameters: (i) the control field amplitude $a_k$, (ii) the control field frequency $\omega_k$, and (iii) the control field phase $\phi_k$. 
Thus, the control field can be described as a function of these three parameters acting on a control Hamiltonian operator $\hat{\sigma}_k$ on the target qubits,
\begin{equation}
    \hat{H}_k = f\left ( a_k (t), \omega_k (t), \phi_k (t)\right )\hat{\sigma}_k = u_k(t) \hat{\sigma}_k
\end{equation}

Adjusting these three parameters over time is the operational definition of pulse control.
We will use the shorthand $u_k(t)$ to denote these three variable functions to generalize over control formalisms for other qubit modalities with different control parameters.
Typically, only the amplitude $a_k (t)$ is tuned to achieve the required control, with the other tunable parameters kept fixed.

\subsection{A classification of quantum optimal control techniques} \label{sec:techniques}

The collection of techniques focused on achieving high-quality control of quantum systems is referred to as quantum optimal control~(QOC). 
It involves methods to design and implement electromagnetic field configurations that can effectively steer quantum processes at the atomic or molecular scale in the best way possible~\cite{koch2022quantum,glaser2015training}. 

QOC techniques can be broadly classified as analytical and numerical methods~\cite{msMD-DE_s_GRAPE_b_GRAPE}. 
As the names suggest, the analytical methods use mathematical theory and representations of the quantum system to analytically solve for the optimal pulse, while numerical methods leverage the power of discretization and linearization to allow the use of numerical methods and algorithms. 
An example of an analytical method for QOC is Pontryagin's maximum principle~\cite{pontryagin}. 
It models any optimal control together with the optimal state trajectory as a 2-point boundary value problem with a maximum condition of the control Hamiltonian. 
Using this description, one can use a time-varying Lagrangian description and multiplier vector to solve the problem. 
Another method to analytically devise optimal control pulses is through a generalization of adiabatic evolution, for example, via derivative removal by adiabatic gates (DRAG) ~\cite{drag}.
If the system is more complex, such as multiple qubits, interactions, or noise systems, often a perturbative expansion~\cite{perturbativeexpansion} is needed to solve it analytically. 
SU(2) Lie algebra~\cite{lierank} can also be used to devise rules analytically when a quantum system is fully reachable or controllable.

\begin{table}[htb]
\centering
\caption{Overview of some existing methods of Quantum Optimal Control, including their category based on analytical, numerical, closed loop, open loop, gradient-based, and gradient-free methods.}
\resizebox{0.9\textwidth}{!}{%
\begin{tabular}{lcccccc}
    \textbf{Quantum Optimal Control method} & \textbf{\rot[35][3em]{Analytical}} & \textbf{\rot[35][3em]{Numerical}} & \textbf{\rot[35][3em]{Closed loop}} & \textbf{\rot[35][3em]{Open loop}} & \textbf{\rot[35][3em]{Gradient-based}} & \textbf{\rot[35][3em]{Gradient-free}}\\
    \\[-1.3em]
    \hline
    Pontryagin's maximum principle & $\checkmark$ &  &  & $\checkmark$  & $\checkmark$ &  \\
    Derivative removal by adiabatic gates & $\checkmark$ &  &  & $\checkmark$  &  & $\checkmark$  \\
    Perturbative expansion & $\checkmark$ &  &  & $\checkmark$  & $\checkmark$ &   \\
    SU(2) Lie algebra & $\checkmark$ &  &  & $\checkmark$ &  & $\checkmark$  \\
    Reinforcement learning &  & $\checkmark$ & $\checkmark$ &  &  &  \\
    Q-Learning &  & $\checkmark$ & $\checkmark$ &  &  &  \\
    msMS-DE &  & $\checkmark$ & $\checkmark$ &  &  &  \\
    Sampling-based learning control &  & $\checkmark$ & $\checkmark$ &  &  &  \\
    s-GRAPE &  & $\checkmark$ & $\checkmark$ &  &  &  \\
    b-GRAPE &  & $\checkmark$ & $\checkmark$ &  &  &  \\
    Krotov &  & $\checkmark$ &  & $\checkmark$ & $\checkmark$ &  \\
    Gradient optimization of analytical control &  & $\checkmark$ &  & $\checkmark$ & $\checkmark$ &  \\
    Gradient ascent pulse engineering &  & $\checkmark$ &  & $\checkmark$ & $\checkmark$ &  \\
    Chopped random basis optimization &  & $\checkmark$ &  & $\checkmark$ &  & $\checkmark$ \\
    Genetic algorithm &  & $\checkmark$ &  & $\checkmark$ &  & $\checkmark$ \\
    Differential evolution &  & $\checkmark$ &  & $\checkmark$ &  & $\checkmark$
    
\end{tabular}%
}
    \label{table:methods}
\end{table}

While analytical methods are, by default, open-loop, numerical methods can further be classified into closed-loop and open-loop methods.
Closed-loop methods do not require an existing theoretical model of the system and use experimental feedback loops, with measurement data from the quantum system, to devise and refine the control pulses.
Contrarily, open-loop methods rely on using an existing theoretical model of the quantum system and process in question and, accordingly, design optimal control pulses based on that theoretical model that is directly applied to the system. 

A popular closed-loop method is a sub-class of machine learning~(ML) called reinforcement learning~(RL). 
In RL, a so-called agent is allowed to take certain actions and apply them to the environment. The environment then outputs a certain state, accompanied by a certain reward based on the action that was taken. 
In framing QOC as an RL problem, the action is the control pulse, the environment is the quantum system, and the state is the output state of the quantum system after applying that specific control pulse~\cite{reinforcement_learning}.
The negative reward (or penalty) is modeled as a measure (e.g., process distance) between the target environment state and the achieved environment state.  
In the case of a model-free RL algorithm, the policy to decide the action at every step can either be updated using various methods like the use of a neural network~(NN) or via a method called Q-Learning~\cite{q-learning}. 
To increase the robustness in the feedback loop, the information of the Hessian matrix can be potentially used in combination with a closed-loop learning-based algorithm, e.g., the msMS-DE algorithm~\cite{msMD-DE_s_GRAPE_b_GRAPE}. 
Samples of a quantum system for training purposes can also be used to evaluate the performance within a test and evaluation phase. 
This is called sampling-based learning control~(SLC)~\cite{sampling_based}. 
Finally, there are hybrid methods that use the gradient-based GRAPE algorithm in combination with reinforcement learning or other machine learning methods, such as s-GRAPE or b-GRAPE~\cite{msMD-DE_s_GRAPE_b_GRAPE}. 

Open-loop methods (analytical and numerical) can further be categorized into gradient-based or gradient-free.
Gradient-based methods rely on the calculation of local gradients to, in return, move towards a local optimum, while gradient-free methods usually use variants of stochastic search algorithms to reach an optimum~\cite{msMD-DE_s_GRAPE_b_GRAPE}. 

An example of a gradient-based open-loop numerical method is the so-called Krotov method~\cite{krotov}. 
This method utilizes Lagrange multipliers based on the process fidelity to find optimal pulses for gate synthesis and state-to-state transfer. 
If representing the control pulse in an analytical form is preferred, we may use the so-called gradient optimization of analytical controls~(GOAT) method~\cite{GOAT}. 
This method uses an educated guess of the shape of the pulse (e.g., Gaussian), to form a coupled system of equations, which can then be solved numerically by forward integration methods, such as the Runge-Kutta method. 
Potentially the most well-known and widely used method is the gradient ascent pulse engineering~(GRAPE) method~\cite{khaneja2005grape}. 
This method uses the discretization of the control operators to iteratively solve for the optimal control pulse. 

An example of gradient-free open-loop numerical methods is the chopped random basis optimization~(CRAB)~\cite{crab}. 
This method leverages the fact that optimal solutions could reside in a low-dimensional subspace of the total search space. 
The control sequences are represented as a linear addition of basis functions. 
Another gradient-free method is based on an evolutionary process called the genetic algorithm~(GA)~\cite{genetic_algorithm}. 
This method utilizes a set of initial random guesses and then evolves these candidates based on a fitness function. 
A variation of GA is the differential evolution~(DE) method~\cite{differential_evolution}.
Table \ref{table:methods} provides an overview of the quantum optimal control technique classifications as discussed. 

The methods surveyed above focus on achieving a high fidelity of control as the parameter defining optimality.
In this research, we use both an open-loop and a closed-loop method to assess whether it is possible to incorporate energy-efficient control pulses that can trade-off an optimality cost accounting for both fidelity and energy. 
To this end, we have chosen the naive gradient ascent pulse engineering~(GRAPE) as the open-loop method and deep reinforcement learning~(DRL) as the closed-loop method that we extend to include energetic cost tuning. 
The motivation for this choice will be discussed in the corresponding sections detailing the implementations. 

\subsection{Energetic cost of a unitary operation} \label{sec:eoc_theory}

The energy of a quantum state is defined as the expectation value of the Hamiltonian,
\begin{equation}
    \langle E \rangle =\langle \psi | \hat{H} |\psi\rangle 
    \label{eq:energy}
\end{equation}

However, the energetic cost of a unitary operation is harder to define.
\cite{deffner2021energeticcost} defines the energetic cost as the time-integrated norm of the Hamiltonian over the total duration of the unitary gate,
\begin{equation}
    E [U] = \int_0^\tau dt \| \hat{H}(t)\|
    \label{eq:ec}
\end{equation}

\begin{figure}[htb]
    \centering 
    \includegraphics[clip, trim=0.9cm 1.1cm 0.7cm 0.5cm, width = 0.25\linewidth]{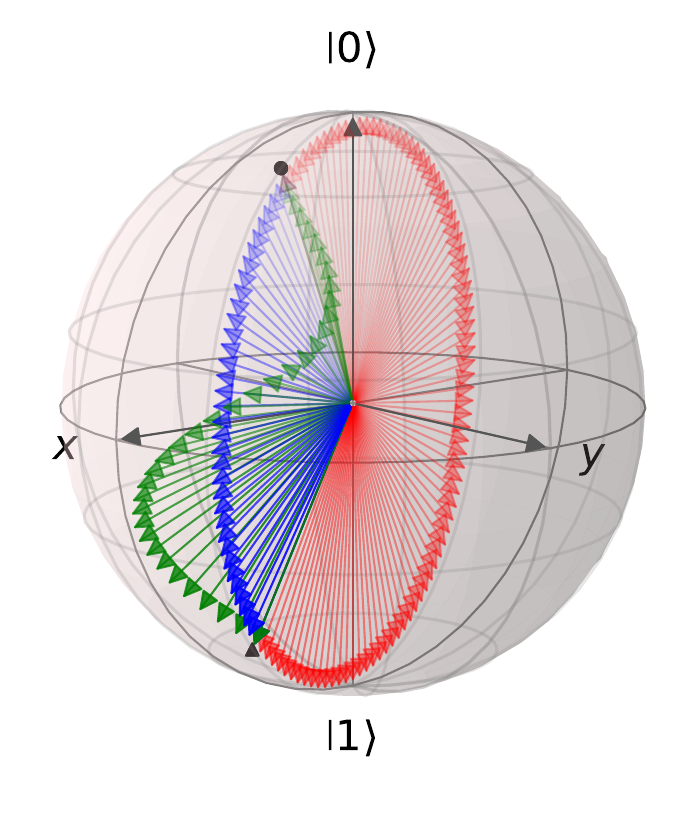}
    \caption{Visual interpretation of an energy-efficient quantum unitary gate. The three paths (blue, red, and green) accomplish the same transformation between the two states.  The optimal geodesic path (in blue) requires a rotation against a single axis along an angle $\le \pi$.}    \label{fig:geodesic}
\end{figure}

For single qubit quantum unitary gates, this cost can be visually interpreted on the Bloch Sphere. 
State-to-state transfer between two quantum states can be accomplished via various paths in Hilbert space.
The most energy-efficient way to implement a single unitary gate corresponds to the geodesic between the two quantum states on the Bloch sphere, i.e., the shorter the path length between the two states, the lower the energetic cost of the quantum unitary gate~\cite{aifer2022quantum}. 
As an example, figure~\ref{fig:geodesic} visualizes different path lengths between two quantum states on the Bloch sphere. 
The unitary gate along the path depicted with blue vectors, in this case, has a lower energetic cost than the unitary gate that implements the red or green vectors. 

In this research, we will investigate the cost governed by equation~\ref{eq:ec}, as well as the path length representation, for computing the energetic cost of a quantum unitary gate. 
Although one cannot visualize multi-qubit unitary gates on the Bloch sphere, similar geometric arguments in corresponding higher dimensional manifolds hold there as well. 

\section{Energy-optimized gradient ascent pulse engineering} \label{sec:eogrape}

In this section, the gradient ascent pulse engineering~(GRAPE) algorithm is reviewed, which uses a gradient that is based on the fidelity of the unitary synthesis.
Thereafter, we derive the gradient of the energetic cost that is used for co-optimizing within our proposed energy-optimized GRAPE algorithm.
Based on this gradient, we present our proposed EO-GRAPE quantum control algorithm.
Our implementation of the algorithm is used to study the resultant control pulses, the trade-off between energy and fidelity of the synthesized target unitary, and the dependence of the algorithm on system noise using a QPU simulator.

\subsection{GRAPE algorithm preliminaries}

The total Hamiltonian of a closed quantum system consists of a constant drift Hamiltonian $\hat{H}_D$, and the sum of control Hamiltonians $\hat{H}_k$ governed by parameters $u_k(t)$. 
Thus, the total system Hamiltonian can be written as,
\begin{equation}
    \hat{H} = \hat{H}_D + \hat{H}_C = \hat{H}_D + \sum_{k=1}^{K}\hat{H}_k = \hat{H}_D + \sum_{k=1}^{K}u_k(t)\hat{\sigma}_k
\end{equation}

The dynamics of a state $\left | \psi (t) \right >$, represented as a vector in a Hilbert space, evolves according to the time-dependent Schrödinger equation as reviewed earlier in equation~\ref{eq:TDSE}.
The time-evolution of the quantum state over time $T$ is discretized into $N$ steps of $\Delta t$.
During a specific time step, $n$, the evolution is given by the unitary operator,
\begin{equation}
    U_n = \exp{\left \{ -i \Delta t  \left ( \hat{H}_D + \sum_{k=1}^K u_k(n)\hat{\sigma}_k\right)\right \}}
\end{equation}

Thus, the total unitary that is implemented after time $T = N \cdot \Delta t$ is given by:
\begin{equation}
    U(t=T) = U_{N} \dots U_{1} = \prod_{n=1}^{N}\exp{\left \{ -i \Delta t  \left ( \hat{H}_D + \sum_{k=1}^K u_k(n)\hat{\sigma}_k\right)\right \}}
\end{equation}
Note that we use the product notation to denote the right-to-left product ordering of the elements.

For a specific time slice, the forward and backward propagators refer to the rest of the unitary after and before this time slice, respectively, such that the total unitary is constructed from their composition. 
The forward propagator $X_n$ is given by,
\begin{equation}
    X_n \equiv U_{n} \dots U_{1}
\end{equation}
Accordingly, the backward propagator $P_n$ can be defined as,
\begin{equation}
    P_n \equiv U_{n+1}^{\dagger} \dots U_{N}^{\dagger}
\end{equation}

We also need to define a performance function that is to be maximized. 
In the case of quantum unitary gate synthesis, this function is defined as the overlap between the target unitary $U_T$ and the final unitary after time $T$,
\begin{equation}
    \Phi = \left | \langle U_T | U(T) \rangle \right |^{2}
\end{equation}

The GRAPE algorithm iteratively updates each control parameter $u_k$ for each time step $n$ according to the gradient of the performance function $\Phi$ with respect to the control parameters $u_k$. 
This is repeated for a certain amount of iterations, called the GRAPE-iterations, and denoted by $N_G$.
This update rule, for an arbitrary step size $\epsilon$, is defined \cite{khaneja2005grape} as,
\begin{equation}
    u_k(n) \rightarrow u_k(n) + \epsilon \frac{\partial \Phi}{\partial u_k(n)}
    \label{eq:update}
\end{equation}

To evaluate the gradient, the derivative of the performance function with respect to the control parameters $u_k(n)$ needs to be evaluated.
To derive this, first, the performance function is expressed in terms of the forward and backward propagators,
\begin{equation}
    \Phi = \langle U_T | U_N \dots U_1 \rangle \langle U_1 \dots U_N | U_T\rangle =  \langle P_n|X_n \rangle \langle X_n | P_n \rangle 
\end{equation}

By using perturbation theory~\cite{khaneja2005grape} to the first order in $\partial u_k(n)$,
\begin{equation}
    \frac{\partial \Phi}{\partial u_k(n)} = -\langle P_n|X_n\rangle \langle i \Delta t \hat{\sigma}_k X_n | P_n\rangle - \langle P_n | i \Delta t \hat{\sigma}_k X_n \rangle \langle X_n | P_n\rangle = -2 \mathscr{Re} \left \{  \langle P_n | i \Delta t \hat{\sigma}_k X_n \rangle \langle X_n | P_n \rangle \right \}
    \label{eq:d-fid}
\end{equation}

By using the definition of the gradient, the forward and backward operator, and by updating the control parameters according to equation~\ref{eq:update}, we can arrive at global maxima of the performance function for a specific set of control parameters $u_k(n)$. 

\subsection{Energy gradient} \label{sec:energy_grad}

In this research, a novel performance function is introduced and utilized, consisting of a fidelity part and an energetic cost part. 

The energetic cost of a discretized pulse based on equation~\ref{eq:ec} is defined as,
\begin{equation}
    \hat{C}[U(T)] = \frac{\sum_{n=1}^{N} \Delta t \left \| \hat{H}_D + \sum_{k=1}^{K} u_k(n)\hat{\sigma}_k \right \|}{T\left \| \hat{H}_D + \sum_{k=1}^{K} \hat{\sigma}_k \right \|} \equiv \phi_e
\end{equation}
For brevity, the normalization constant in the denominator will be referred to as $N_e$.


We define the total cost function as,
\begin{equation}
    \Phi = w_{f} \left ( 1 - \left | \frac{\mathrm{Tr}(U_{T}^{\dagger} U(T)}{\mathrm{Tr}(U_{T}^{\dagger} U_{T})} \right |^{2} \right ) + w_{e} \left( \frac{1}{N_e} \sum_{n=1}^{N} \Delta t \left \| \hat{H}_D + \sum_{k=1}^{K} u_k(n)\hat{\sigma}_k \right \| \right )
\end{equation}

Thus, the gradient of $\Phi$ with respect to $u_k(n)$ can be written as the weighted linear addition of the two components of the cost function based on fidelity, $\phi_f$ and energy, $\phi_e$,
\begin{equation}
    \frac{\partial \Phi}{\partial u_k(n)} = w_{f} \frac{\partial \phi_f}{\partial u_k(n)} + w_{e} \frac{\partial \phi_e}{\partial u_k(n)}
\end{equation}

The gradient of fidelity $\phi_f$ based on equation~\ref{eq:d-fid} is,
\begin{equation}
    \frac{\partial \phi_f}{\partial u_k(n)} = -2 \mathscr{Re} \left \{  \langle P_n | i \Delta t \hat{\sigma}_k X_n \rangle \langle X_n | P_n \rangle \right \}
    \label{eq:totalphif}
\end{equation}

The energetic cost part of the cost function can be written as:

\begin{equation}
    \phi_e = \frac{1}{N_e} \sum_{n=1}^{N} \Delta t \left \| \hat{H}_D + \sum_{k=1}^{K} u_k(n)\hat{\sigma}_k \right \| 
\end{equation}

Using the identity $\|A\|=\sqrt{\operatorname{Tr}\left(A^* A\right)}$, the expression above can be expanded as follows:

\begin{equation}
    \phi_e=\frac{1}{N_{e}}\sum_{n=1}^N \Delta t \sqrt{\operatorname{Tr}\left(\hat{H}_D^* \hat{H}_D+\sum_{k=1}^K u_k(n)\left(\hat{H}_D^* \hat{\sigma}_k+\hat{\sigma}_k^* \hat{H}_D\right)+\sum_{k=1}^K \sum_{k^{\prime}=1}^K u_k(n) u_{k^{\prime}}(n) \hat{\sigma}_k^* \hat{\sigma}_{k^{\prime}}\right)}
\end{equation}

The gradient can then be expressed in terms of $f(g(u_k(n)))$ and $g(u_k(n))$:
\begin{equation}
    \frac{\partial \phi_e}{\partial u_k(n)}=\frac{\partial f}{\partial g} \frac{\partial g}{\partial u_k(n)}
\end{equation}
The first part is given by:
\begin{equation}
    \frac{\partial f}{\partial g}=\frac{1}{N_{e}}\sum_{n=1}^N \Delta t \frac{1}{2 \sqrt{g\left(u_k(n)\right)}}
\end{equation}
while the second part is given by:
\begin{equation}
    \frac{\partial g}{\partial u_k(n)}=\operatorname{Tr}\left(\sum_{k=1}^K \hat{H}_D^* \hat{\sigma}_k+\hat{\sigma}_k^* \hat{H}_D\right)+\operatorname{Tr}\left(\sum_{k=1}^K \sum_{k'=1}^K \left(u_k(n)+u_{k^{\prime}}(n)\right)\hat{\sigma}_k^* \hat{\sigma}_{k^{\prime}}\right) 
\end{equation}

The total gradient of $\phi_e$ with respect to $u_k(j)$ can thus be written as:

\begin{equation}
    \frac{\partial \phi_e}{\partial u_k(n)} = \frac{1}{N_e}\sum_{n=1}^N \Delta t \frac{\operatorname{Tr}\left(\sum_{k=1}^K \hat{H}_D^* \hat{\sigma}_k+\hat{\sigma}_k^* \hat{H}_D\right)+\operatorname{Tr}\left(\sum_{k=1}^K \sum_{k'=1}^K \left(u_k(n)+u_{k^{\prime}}(n)\right)\hat{\sigma}_k^* \hat{\sigma}_{k^{\prime}}\right)}{2 \sqrt{\operatorname{Tr}\left(\hat{H}_D^* \hat{H}_D+\sum_{k=1}^K u_k(n)\left(\hat{H}_D^* \hat{\sigma}_k+\hat{\sigma}_k^* \hat{H}_D\right)+\sum_{k=1}^K \sum_{k^{\prime}=1}^K u_k(n) u_{k^{\prime}}(n) \hat{\sigma}_k^* \hat{\sigma}_{k^{\prime}}\right)}}
    \label{eq:totalphie}
\end{equation}

\subsection{EO-GRAPE algorithm} \label{sec:eogrape-algo}

Combining equation~\ref{eq:totalphif} with equation~\ref{eq:totalphie}, the total gradient of $\Phi$ with respect to the control parameters $u_k(n)$ is given as,
\begin{equation}
    \frac{\partial \Phi}{\partial u_k(n)} = \begin{pmatrix}
-2w_f
 \\
-w_e
\end{pmatrix}'
\begin{pmatrix}
\mathscr{Re} \left \{  \langle P_n | i \Delta t \hat{\sigma}_k X_n \rangle  \langle X_n | P_n \rangle \right \}
 \\
\frac{1}{N_e}\sum_{n=1}^N \Delta t \frac{\operatorname{Tr}\left(\sum_{k=1}^K \hat{H}_D^* \hat{\sigma}_k+\hat{\sigma}_k^* \hat{H}_D\right)+\operatorname{Tr}\left(\sum_{k=1}^K \sum_{k'=1}^K \left(u_k(n)+u_{k^{\prime}}(n)\right)\hat{\sigma}_k^* \hat{\sigma}_{k^{\prime}}\right)}{2 \sqrt{\operatorname{Tr}\left(\hat{H}_D^* \hat{H}_D+\sum_{k=1}^K u_k(n)\left(\hat{H}_D^* \hat{\sigma}_k+\hat{\sigma}_k^* \hat{H}_D\right)+\sum_{k=1}^K \sum_{k^{\prime}=1}^K u_k(n) u_{k^{\prime}}(j) \hat{\sigma}_k^* \hat{\sigma}_{k^{\prime}}\right)}}
\end{pmatrix}
\label{eq:totalfullgradient}
\end{equation}

Based on these gradients that were derived in section~\ref{sec:energy_grad} leading to  equation~\ref{eq:totalfullgradient}, we can now formulate our novel EO-GRAPE algorithm that employs the updating rule of,
\begin{equation}
    u_k(n) \rightarrow \epsilon_f \frac{\partial \phi_f}{\partial u_k(n)}+\epsilon_e\frac{\partial \phi_e}{\partial u_k(n)}
    \label{eq:updaterule}
\end{equation}

The EO-GRAPE can be implemented as algorithm \ref{alg:eogrape}.

\begin{algorithm}
\caption{Energy-Optimized Gradient Ascent Pulse Engineering}\label{alg:eogrape}
\begin{algorithmic}
\item $\{u_k(n)\}$ = Initial guess of $u_k(n)$ for all $k$ 

\For{g = $1$ to $N_G$}
\For{n = $1$ to $N$}
\For{k= $1$ to $K$}
\State Calculate $X_n$
\State Calculate $P_n$
\State Evaluate $\partial \phi_f/\partial u_k(n)$ 
\State Evaluate $\partial \phi_e/\partial u_k(n)$
\State Update control amplitudes $u_k(n) \rightarrow \epsilon_f \frac{\partial \phi_f}{\partial u_k(n)}+\epsilon_e\frac{\partial \phi_e}{\partial u_k(n)}$
\EndFor
\EndFor
\EndFor
\State \Return $\{u_k(n)\}$
\end{algorithmic}
\end{algorithm}

To test the fidelity-energy trade-off and for the other experiments presented later in this research, we consider two exemplary drift Hamiltonians:
\begin{equation}
    \hat{H}_D^{1} \in \Bigg\{ \hat{\sigma}_i, \frac{\hbar \omega_1}{2} \hat{\sigma}_z \Bigg\}
    \quad \hat{H}_D^{2} \in \Bigg\{ \frac{\hbar \omega_1}{2} \hat{\sigma}_z^{(1)} \otimes \hat{I}_2^{(2)} + \frac{\hbar \omega_2}{2} \hat{I}_2^{(1)} \otimes \hat{\sigma}_z^{(2)} + \hbar J \hat{\sigma}_z^{(1)} \otimes \hat{\sigma}_z^{(2)} \Bigg\}
\end{equation}
Where $\omega$ is the eigenfrequency of the target qubit, and $J$ is the coupling strength between the two target qubits. 

\begin{figure}[!ht]
     \centering
     \begin{subfigure}[b]{0.48\textwidth}
         \centering
         \includegraphics[width=\textwidth]{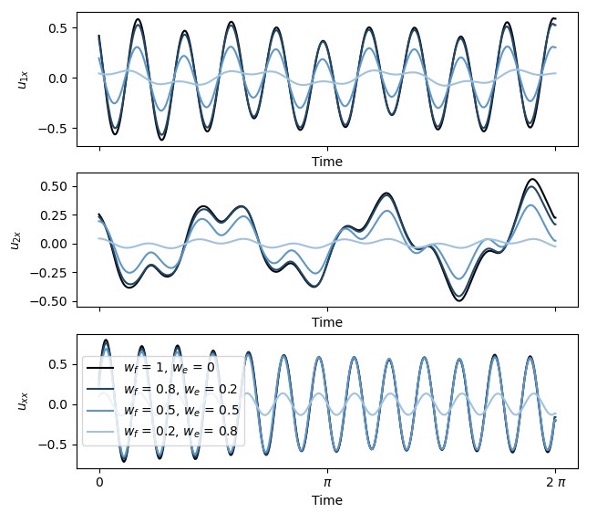}
         \label{fig:randompulse1}
     \end{subfigure}
     \begin{subfigure}[b]{0.485\textwidth}
         \centering
         \includegraphics[width=\textwidth]{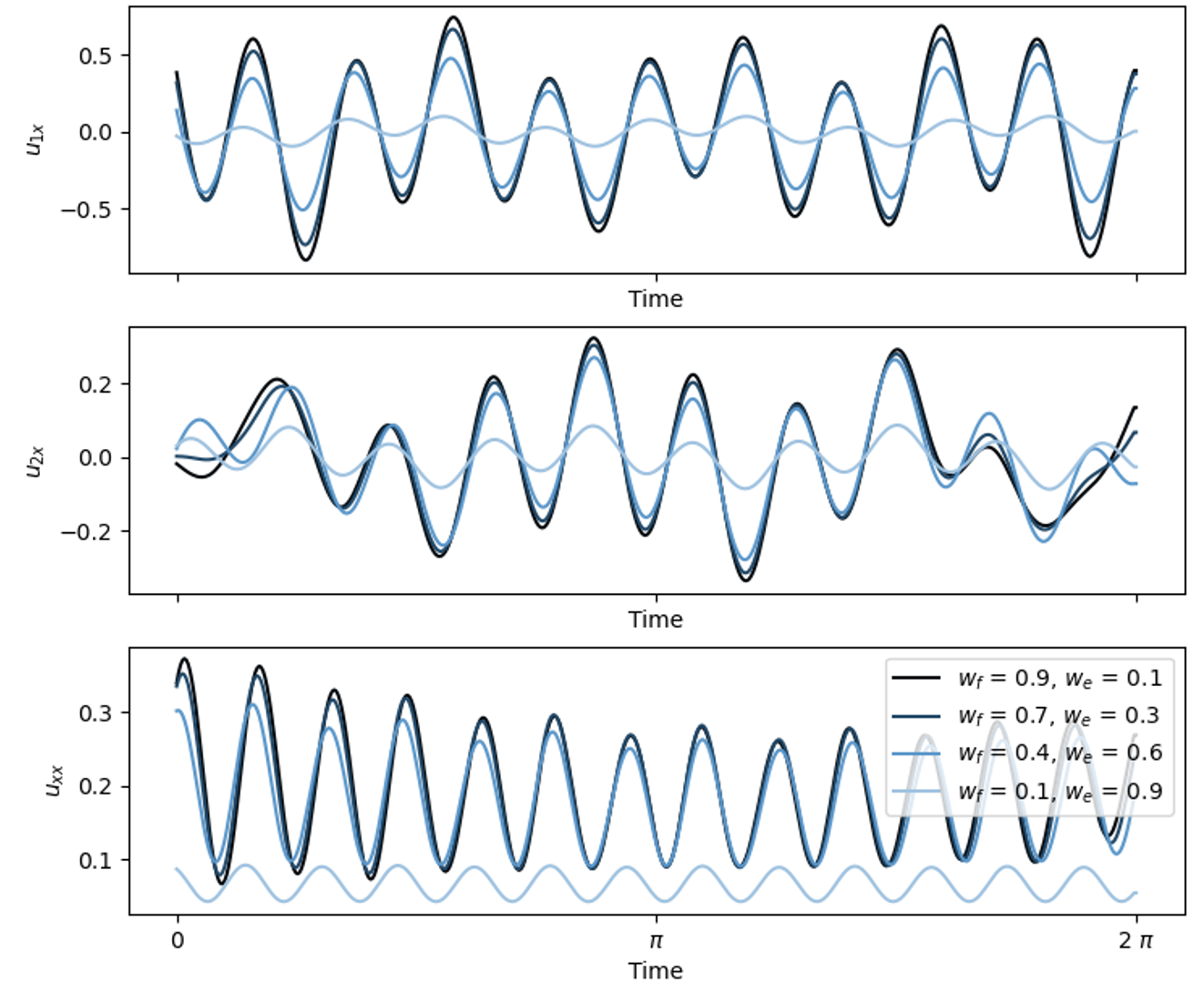}
         \label{fig:randompulse2}
     \end{subfigure}
    \caption{Example control pulses generated by the EO-GRAPE algorithm for different weight settings. Parameters: $U_T = $ \textit{RAND}, $\hat{H}_D = \hat{H}_D^{2}$, $\hat{\sigma}_k \in \{ \hat{\sigma}_{x}^{1}, \hat{\sigma}_{x}^{2}, \hat{\sigma}_x^{1} \hat{\sigma}_x^{2} \} $, $T_1 = \infty $, $T_2 = \infty$,  $w_f = [1, 0.1]$, $w_e = [0, 0.9]$, $N = 500$, $N_G = 500$}
    \label{fig:randompulse}
\end{figure}

In figure \ref{fig:randompulse}, the effect of increasing or decreasing the weight associated with fidelity $w_f$ and energetic cost $w_e$ is shown. As one can see, the higher the value of $w_e$, the lower the amplitude of the control pulses are, decreasing the area and thus decreasing the energetic cost, intuitively matching our expectations. One can also see some other harmonics being introduced when increasing $w_e$ to decrease the energetic cost of the pulses. An overview of the parameters such as $U_T$, $\hat{H}_D$, $\hat{\sigma}_k$, and others are given in the figure caption.

\subsection{Trade-off between fidelity and energetic cost}

Having derived the theoretical energetic cost and an algorithmic procedure to optimize pulses based on the energy gradient, in this section, we explore the following research question: what is the relation between fidelity and the energetic cost?
Naively, one would expect that if the control amplitudes are zero, i.e., no energetic cost, the drift Hamiltonian will be solely responsible for the evolution.
Thus, we expect a fidelity score that depends on the relation between the target unitary and the unitary implemented by the drift Hamiltonian for the pulse duration.

Note that the trade-off between energetic cost and fidelity has already been studied from the perspective of noise and dissipative dynamics~\cite{van2024fidelity}.
While having similar motivations, these are unrelated to the research presented in this article in that we consider all intermediate states in the state evolution to be pure without any environmental coupling.

\begin{figure}[!tb]
     \centering
     \begin{subfigure}[b]{0.48\textwidth}
         \centering
         \includegraphics[width=\textwidth]{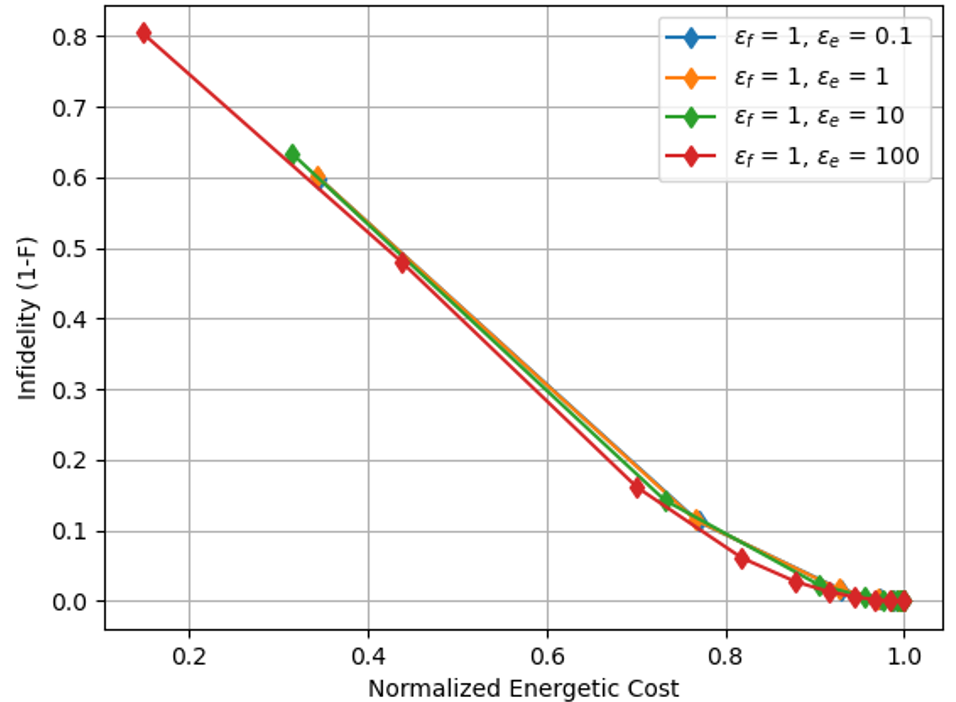}
         \caption{}
         \label{fig:pareto1}
     \end{subfigure}
     \begin{subfigure}[b]{0.48\textwidth}
         \centering
         \includegraphics[width=\textwidth]{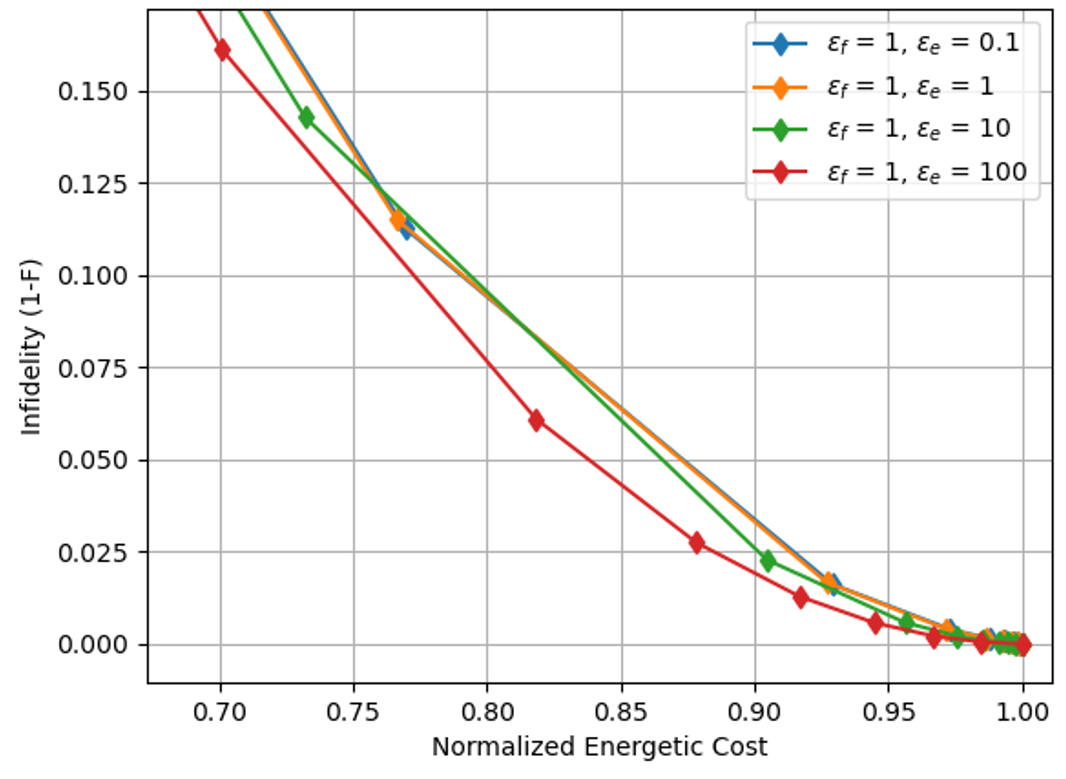}
         \caption{}
         \label{fig:pareto2}
     \end{subfigure}
    \caption{\textbf{(a)} Infidelity and energetic cost values for $10$ different weight settings (indicated by diamond markers for each color) and learning rates $\epsilon_e$, showing the trade-off or Pareto front between fidelity and energetic cost. \textbf{(b)} Zoomed-in view of the Pareto front between fidelity and energetic cost. Parameters: $U_T =$ \textit{RAND}, $\hat{H}_D = \hat{H}_D^{2}$, $\hat{\sigma}_k \in \{ \hat{\sigma}_{x}^{1}, \hat{\sigma}_{x}^{2}, \hat{\sigma}_x^{1} \hat{\sigma}_x^{2} \} $, $T_1 = \infty $, $T_2 = \infty$,  $w_f = [1, 0.1]$, $w_e = [0, 0.9]$, $N = 500$, $N_G = 100$}
    \label{fig:pareto}
\end{figure}

To mitigate the dependence on the drift Hamiltonian, we sample the target unitary over a 2-qubit Harr random distribution.
In figure \ref{fig:pareto}, the results of this experiment are depicted. 
The EO-GRAPE algorithm was run for different sets of weights $w_f$ and $w_e$. 
Each diamond marker on the plot represents a different weight setting between $[w_f:w_e] = [1:0]\dots[0.1:0.9]$ in steps of $0.1$. 
These values have been plotted against each other for different values of the learning rates, $\epsilon_f$ and $\epsilon_e$. 
The Pareto front depicting the trade-off between fidelity and the energetic cost is clearly visible, where a reduction in energetic cost directly leads to an increase in infidelity or a decrease in fidelity.
From our learning rate optimization analysis, it can be concluded that the optimal hyperparameter is $\epsilon_f = 1$ and $\epsilon_e = 100$.
 
\begin{figure}[bh]
    \centering
    \includegraphics[width = 0.5 \linewidth]{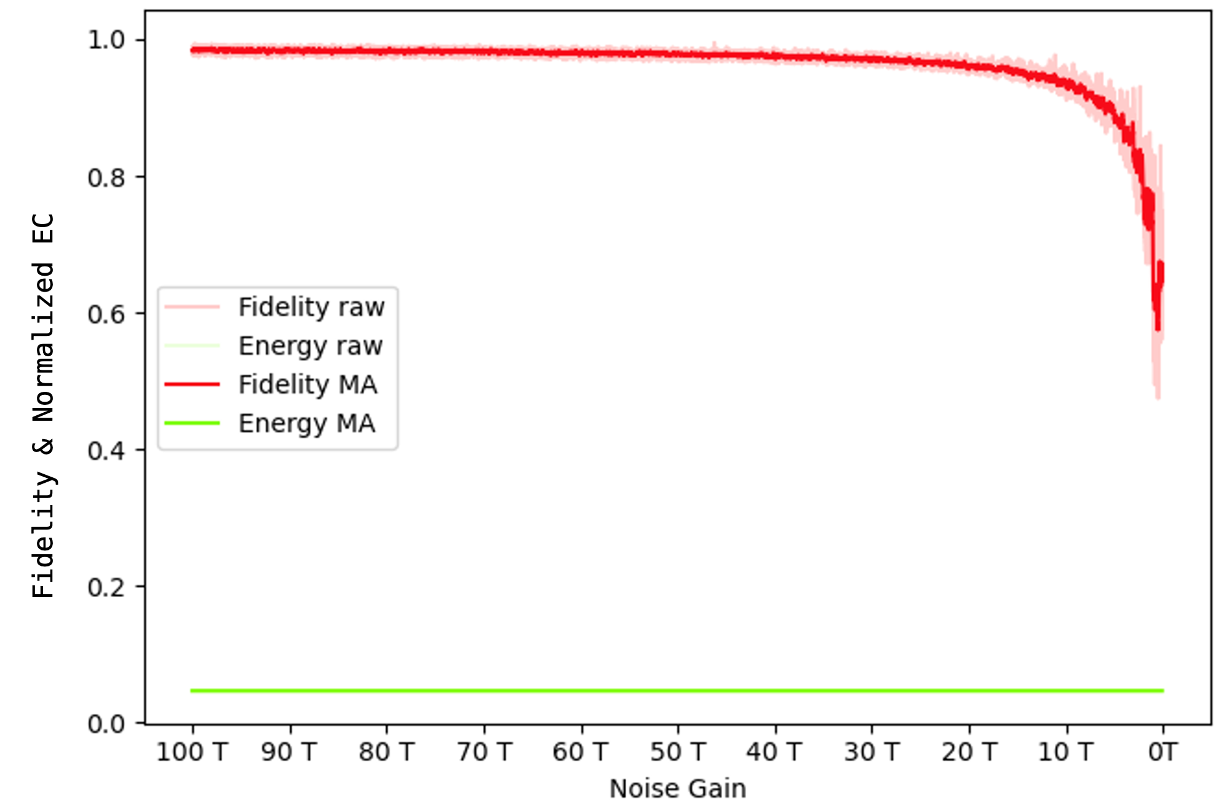}
    \caption{Fidelity (red) and Energetic Cost (green) of EO-GRAPE generated control pulses as a function of decreasing decoherence time, or increasing noise level. The moving average~(MA) is shown in bold colors. Parameters: $U_T =$ \textit{Hadamard}, $\hat{H}_D = \hat{H}_D^{1}$, $\hat{\sigma}_k \in \{ \hat{\sigma}_{x}^{1} \}$, $T_1 = [100T, 1T] $, $T_2 = [100T, 1T]$, $w_f = 0.7$, $w_e = 0.3$, $N = 100$, $N_G = 500$.}
    \label{fig:grapenoise}
\end{figure}

This result matches our intuition that to achieve a lower energetic cost for a quantum unitary gate, one needs to inherently decrease the area or amplitude of the control pulses, resulting in a lower process fidelity.
Nevertheless, as seen on the high fidelity end of the curve in figure~\ref{fig:pareto2}, the two factors are not completely inversely proportional to each other. We can, therefore, still decrease the energetic cost of a quantum unitary gate by roughly 10\% while decreasing the fidelity by roughly 1\%, which can be compensated with error correction~\cite {merrill2014progress}. 
As an estimate from this example of a noise-less setting, one could decrease the energetic cost of each quantum unitary gate by 10\% while maintaining a minimum 2-qubit gate fidelity of 99\%.
Thus, \textit{using EO-GRAPE, we can minimize the energetic cost while maintaining the gate fidelity above the error budget of the quantum application or the threshold of the quantum error correction code}.

To test the performance of the EO-GRAPE algorithm while increasing the noise in the system, we inject an increasing noise over the EO-GRAPE iterations.
From the result of the infidelity, as shown in figure \ref{fig:grapenoise}, we infer that the algorithm is able to achieve high fidelity and low energetic cost throughout most noise settings. 
From a noise gain setting of $20T$ (which is unrealistically high for current quantum devices), we can see that the performance decreases quite rapidly.

\section{Energy-optimized deep reinforcement learning based pulse engineering} \label{sec:eo-drlpe}

As introduced in Section~\ref{sec:techniques}, reinforcement learning can be used for quantum pulse engineering for unitary gate synthesis.
In such a formulation, the action of the RL agent is the control pulse, the environment is the quantum system, and the observation is the output state of the quantum system.
The policy of an RL agent refers to the mechanism to decide the action based on the history of past actions and observations, with additional hyperparameters like the horizon, exploration-exploitation trade-off, etc.
In model-free RL, the policy is often implemented as a neural network~(NN).
The weights of the NN are updated by training on known environments.
In this section, we introduce the agents for energy-optimized deep reinforcement learning-based pulse engineering~(EO-DRLPE).
Thereafter, we present the architectural design of the QOC strategy and its performance in noiseless and noisy cases.

\subsection{Agent design for EO-DRLPE}

In this research, two different reinforcement learning agents are utilized. 
Reinforcement Learning Agent 1 (RLA-1) interacts with the quantum environment and is responsible for actually devising the pulses. 
Reinforcement Learning Agent 2 (RLA-2) is responsible for approximating pulses generated by the EO-GRAPE algorithm.
This can potentially be used as an initial policy of RLA-1. 
If RLA-2 is used, we will refer to this as a warm start~\cite{ash2020warm}, as a pre-training and transfer learning method to initialize the weights of RLA-1 as opposed to random initialization.

\begin{figure}[bht]
    \centering
    \includegraphics[width = 0.7\linewidth]{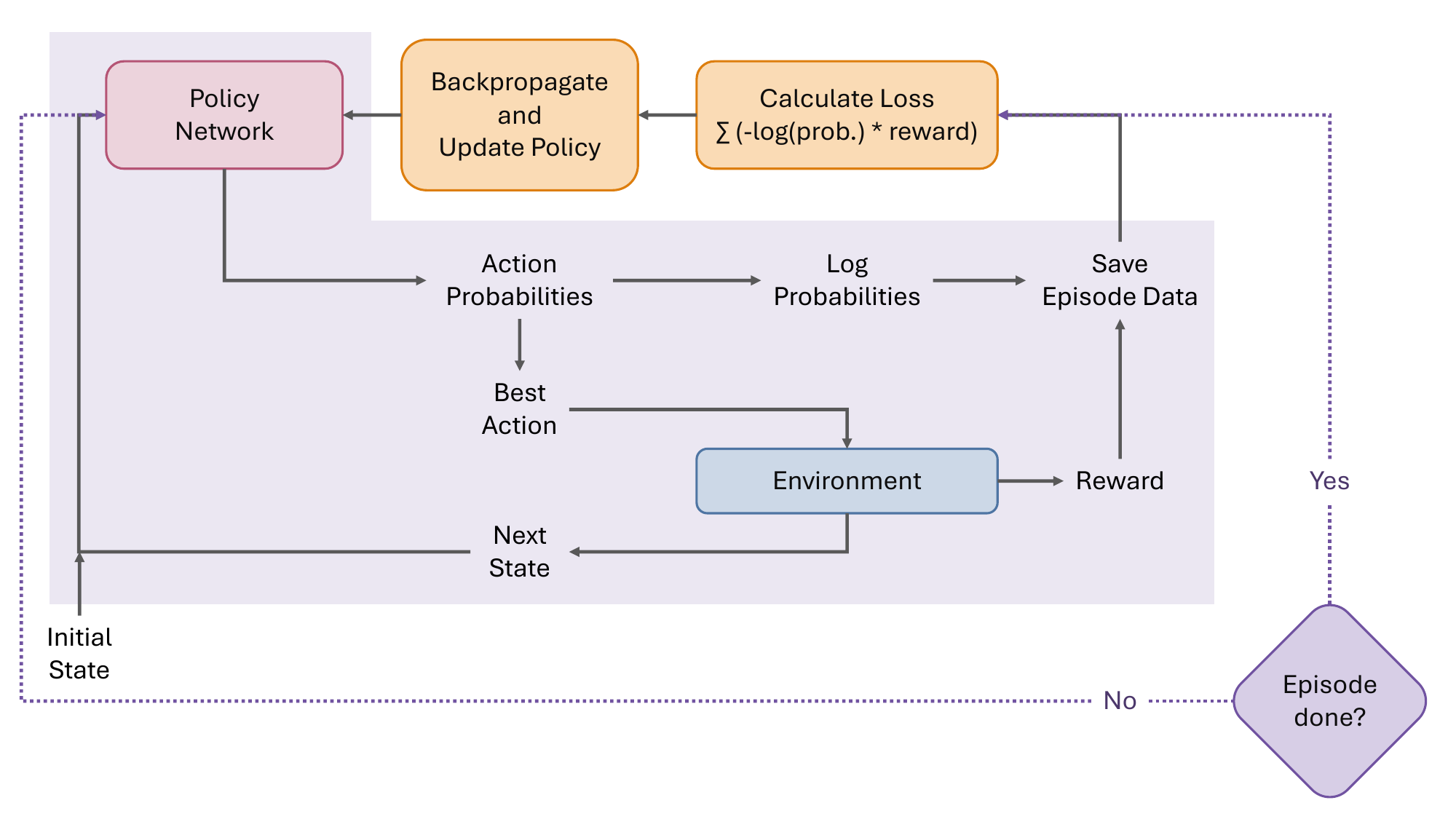}
    \caption{Schematic drawing of the basic workings of the TensorFlow REINFORCE agent \cite{tfreinforce}}
    \label{fig:rl-schematic}
\end{figure}

Both agents use the TensorFlow REINFORCE algorithm~\cite{tensorflow}. 
First, the agent observes the state returned by the environment and chooses an action based on its policy. 
When the action is sent to the environment, the agent will receive a corresponding reward.
The actions and rewards are continuously being registered in the replay buffer. 
After a certain number of actions, the policy is updated based on the use of a deep feed-forward neural network. 
Figure \ref{fig:rl-schematic} shows an overview of the different components and steps in the REINFORCE algorithm\cite{tfreinforce}. 

For RLA-1, the Environment is a quantum simulator from QuTip~\cite{qutip}, called the QuTip Processor class. 
The environment has several input parameters, such as the number of qubits $N_q$, the drift Hamiltonian $\hat{H}_D$, the control Hamiltonians $\hat{H}_k$, and the decoherence times of each qubit $T_1$ and $T_2$. 
The agent is allowed to implement an action, which in this case are the control pulses $u_k(n)$, with shape $(K \times N)$. The output of the environment, called the state, is, in this case, the output density matrix $\rho_{out}$ of the quantum system after applying the action. 
The next decision of the agent is based on the state and the reward, which in our case is defined as,
\begin{equation}
    r_\text{RLA-1} = w_f F(\rho_T, \rho_{out}) + w_e (1 - \hat{C} [U(T)])
\end{equation}

The reward is a linear weighted addition of the fidelity $F$ between the target output density matrix $\rho_T$ and the output density matrix after applying the action $\rho_{out}$, and the inverse of the normalized energetic cost of implementing a certain Unitary after time $T$, $\hat{C}[U(T)]$.

The agent makes a decision based on its policy function $\pi (A_n, S_n)$, which is constantly being updated by a neural network. 
For RLA-1, a deep feed-forward neural network with $1$ input layer, $3-5$ hidden layers, and $1$ output layer is utilized. 

The goal of RLA-2 is to mimic the pulses generated by the EO-GRAPE algorithm by minimizing the distance between the control pulses it generates and the control pulses generated by the EO-GRAPE algorithm as discussed in Section~\ref{sec:eogrape-algo}. 
The agent interacts with a custom Environment class, called the GRAPEApproximation class, that takes in a set of control pulses and outputs the theoretical unitary.
The environment has input parameters such as the number of qubits $N_q$, the drift Hamiltonian $\hat{H}_D$, the control Hamiltonians $\hat{H}_k$, the number of time steps $N$, the total time $T$, and the number of EO-GRAPE iterations $N_G$. 
The agent is allowed to take an action, i.e., the control pulses $u_k(n)$, with shape $(K \times N)$. 
The state that the environment returns to the agent is the theoretical unitary that the control pulses implement based on unitary evolution $U(u_k(n))$. 
RLA-2 utilizes a deep feed-forward NN with a similar configuration as RLA-1.
The RLA-2 learns based on the reward that the environment returns, which is the L1 norm between the target pulse generated by the EO-GRAPE algorithm and the action that the agent takes,
\begin{equation}
    r_\text{RLA-2} = - \sum_{n=1}^{N} \sum_{k=1}^{K} \left | u_{k}^{EO-GRAPE}(n) - u_{k}^{RLA-2}(n) \right |^{2}
    \label{eq:sqdiff}
\end{equation}

\subsection{Workflow of pulse-level unitary synthesis}

\begin{figure}[bht]
    \centering
    \includegraphics[width = 0.65 \linewidth]{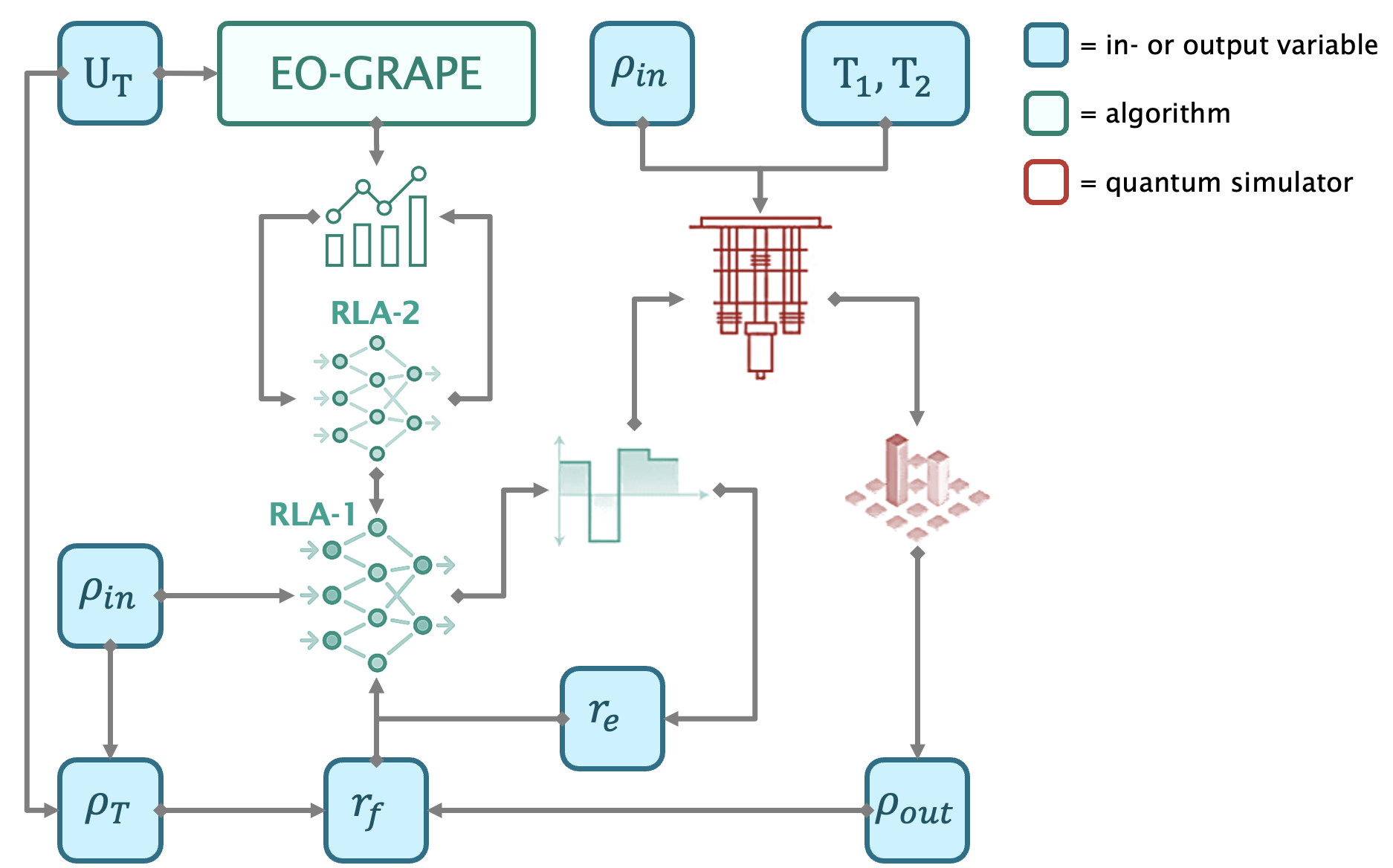}
    \caption{A schematic overview of the different classes and how they are co-dependent on each other. The red icons indicate quantum simulators, the blue icons indicate the input and output variables of the algorithms or simulators, and the green icons indicate an algorithm.}
    \label{fig:codearchitecture}
\end{figure}

Our implementation of the software accompanying this research is contained in a comprehensive open-sourced Python package available on \href{https://github.com/QML-Group/EO-QCtrl}{GitHub}.
The package contains four main classes, with each having a different objective and dependencies.
The four classes include two quantum environment classes and two reinforcement learning classes. The first quantum environment class (depicted by the red quantum simulator in figure \ref{fig:codearchitecture}) interacts directly with the first reinforcement learning class (depicted by the RLA-1 neural network). The second quantum environment class (depicted by the graph output from EO-GRAPE) only interacts with the second reinforcement learning class (depicted by the RLA-2 neural network).
Figure \ref{fig:codearchitecture} shows a schematic overview of the four classes working together on the quantum optimal control problem of unitary pulse synthesis.

\subsection{Performance study}

\begin{figure}[hbt]
    \centering
    \includegraphics[width = 0.5 \linewidth]{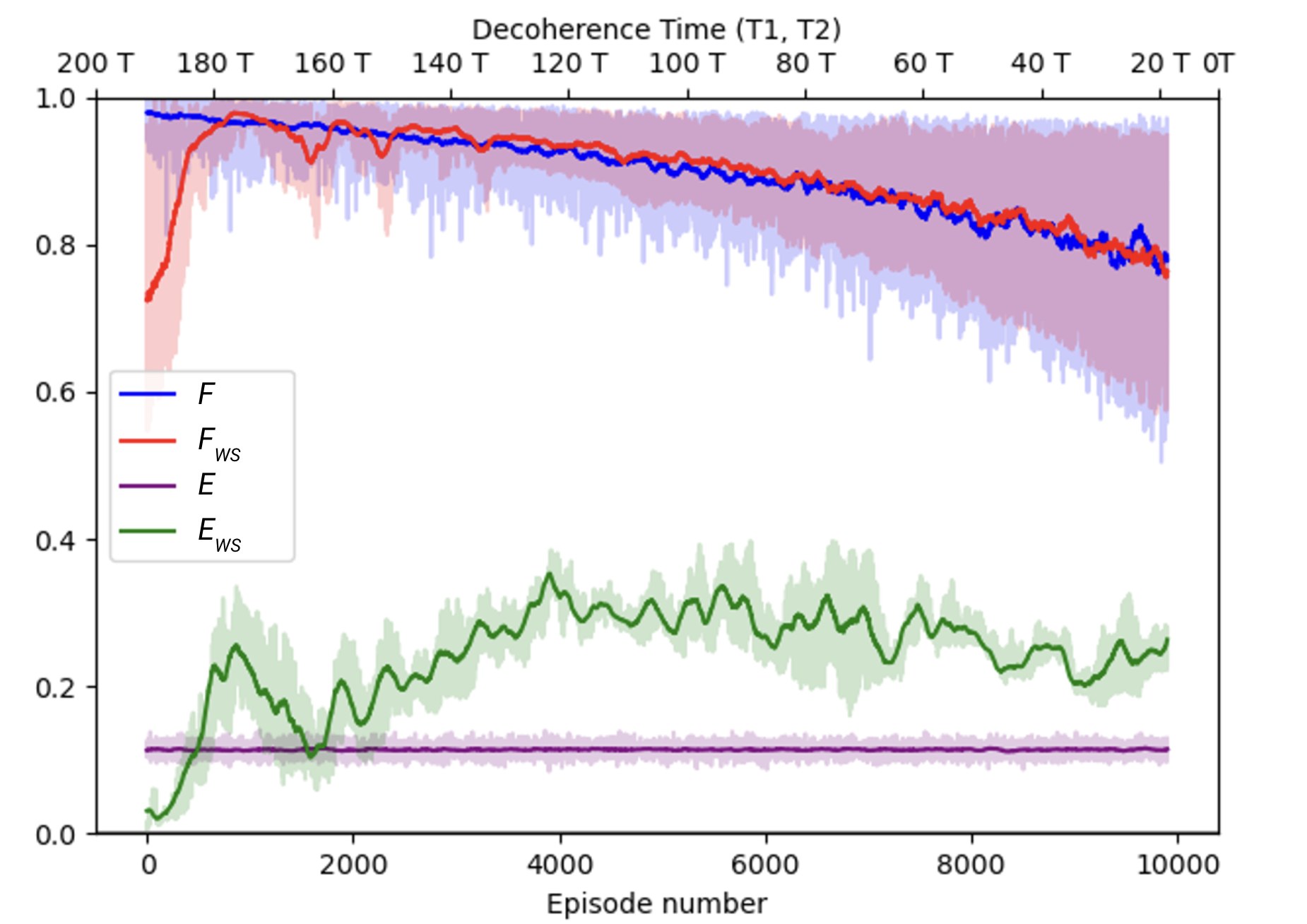}
    \caption{Fidelity (blue) and energetic cost (purple) of RL generated control pulses with warm start, and fidelity (red) and energetic cost (green) of RL generated control pulses without warm start, as a function of the training episode number and decreasing decoherence time. Parameters: $U_T =$ \textit{Hadamard}, $\hat{H}_D = \hat{H}_D^{1}$, $\hat{\sigma}_k \in \{ \hat{\sigma}_{x}^{1} \}$, $T_1 = [200T, 1T] $, $T_2 = [200T, 1T]$, $w_f = 0.8$, $w_e = 0.2$, $N = 100$, $N_G = 500$, \textit{layer\_params} = $(200, 100, 50, 30, 10)$, $N_{QRLA} = 10000$, $N_{GA} = 2000$.}
    \label{fig:rlnoise}
\end{figure}

To evaluate the performance of the EO-DRLPE implementation, we evaluate the fidelity and energetic cost of implementing the Hadamard gate.
The noise in the system is gradually increased (i.e., the $T1$ and $T2$ time is decreased) to assess the robustness of the system in noisy quantum processors.
In figure \ref{fig:rlnoise}, the performance of the reinforcement learning agent both with and without a warm start (by RLA-2) is shown. 
As we can infer, the agent without a warm start (in red) is able to reach slightly higher fidelity than the agent with a warm start (in blue).
However, the energetic cost of the agent with a warm start (in purple) is more stable and lower than a randomly initialized RLA-1 (in green). 

The EO-GRAPE algorithm calculates control parameters based on the target unitary and the drift and control Hamiltonian that are provided and thus does not take into account any noise in the system. 
It is, therefore, quite remarkable that the control pulses generated by the EO-GRAPE algorithm are still able to achieve high fidelity and low energetic costs in a highly noisy environment. 
The reinforcement learning agent without a warm start seems to be able to learn the noise characteristics of the system first and could, therefore, outperform the reinforcement learning agent with a warm start on fidelity. 
However, when we take into account the energetic cost as well, it is clear that the reinforcement learning agent with a warm start outperforms the agent without a warm start. 

\section{Geodesic analysis} \label{sec:geodesic}

In this section, we investigate the effect of the control pulses generated by the EO-GRAPE algorithm and the EO-DRLPE agents on the unitary paths on the Bloch sphere.
As illustrated in figure~\ref{fig:geodesic}, the optimal path on the Bloch sphere is directly proportional to the optimal energetic cost of the unitary operation, conditioned on a fixed total evolution time.
While the relation between geodesic and energetic cost has been previously investigated in its theoretical rigor~\cite{deffner2021energeticcost}, in this research, we explore the pragmatic geodesic achieved by algorithmic procedures of energy-optimized QOC, i.e., EO-GRAPE and EO-DRLPE.  
Co-optimizing the total evolution time guided by quantum speed limits~\cite{aifer2022quantum} along with fidelity and energetic cost for pragmatic quantum pulse engineering is left to future work.

\begin{figure}[!ht]
     \centering
     \begin{subfigure}[b]{0.48\textwidth}
        \centering 
        \includegraphics[clip, trim=0.2cm 0.2cm 0.2cm 0.2cm, width=0.65\textwidth]{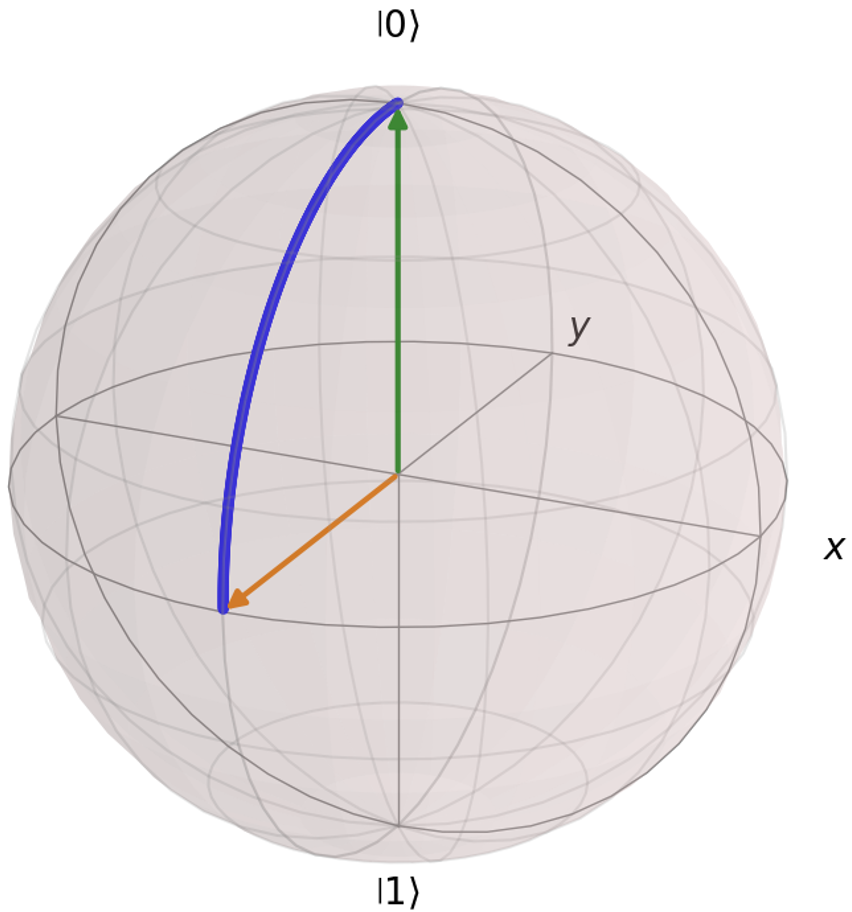}
         \caption{}
         \label{fig:blochidentity1}
     \end{subfigure}
     \begin{subfigure}[b]{0.48\textwidth}
        \centering 
        \includegraphics[clip, trim=0.2cm 0.2cm 0.2cm 0.2cm, width=0.62\textwidth]{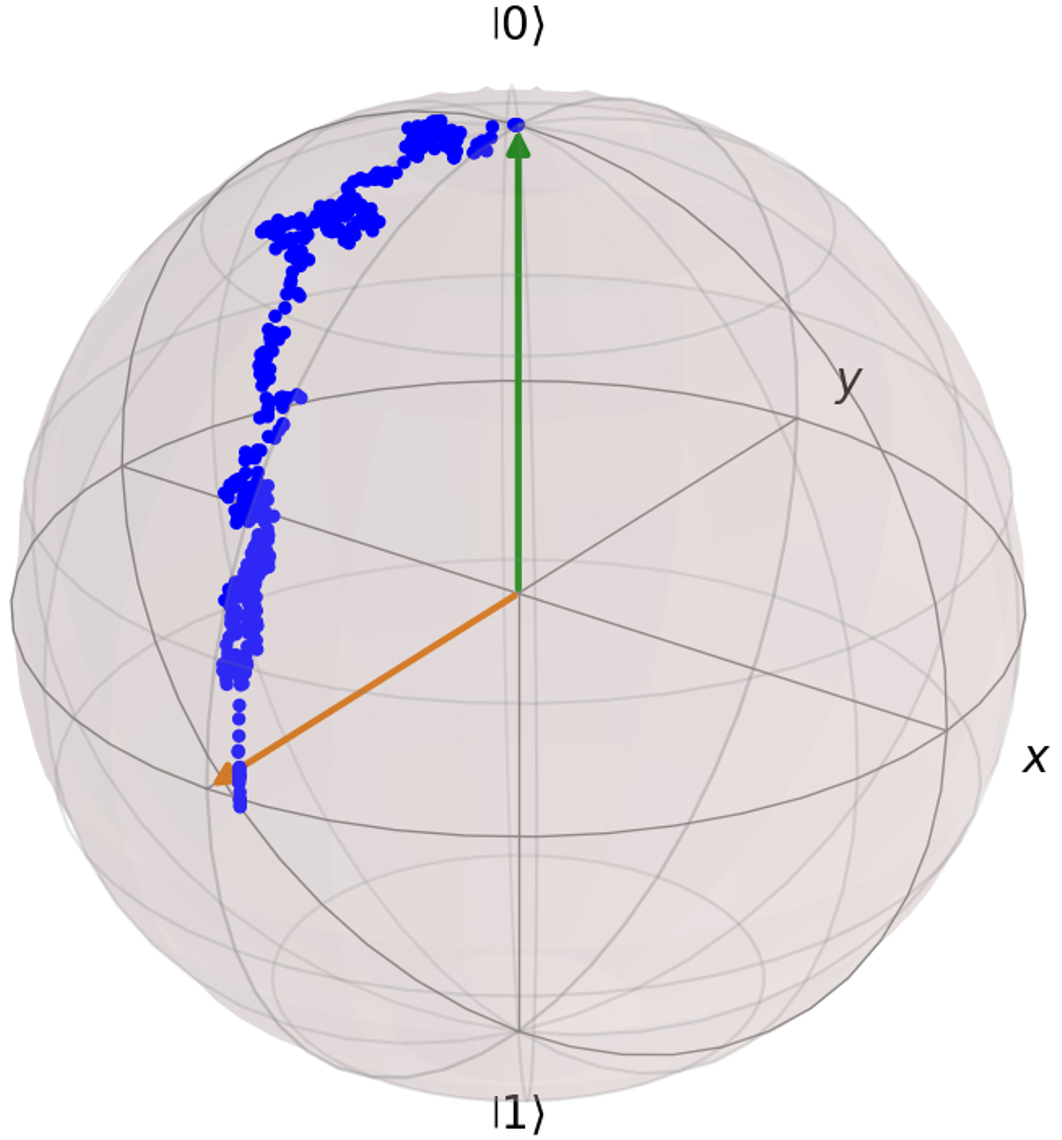}
         \caption{}
         \label{fig:blochidentity2}
     \end{subfigure}
    \caption{Initial state (green vector), target state (orange vector), and path of the quantum unitary (blue) in the absence of a drift Hamiltonian $\hat{H}_D$, by \textbf{(a)} EO-GRAPE generated pulses and \textbf{(b)} RL generated pulses on the Bloch Sphere. Parameters: $U_T = R_{x} (\pi/2)$, $\hat{H}_D = I_2$, $\hat{\sigma}_k \in \{ \hat{\sigma}_{x}^{1}, \hat{\sigma}_{y}^{1} \}$, $T_1 = 1000T$, $T_2 = 1000T$, $w_f = 1$, $w_e = 0$, $N = 500$, $N_G = 500$, $N_{QRLA} = 10000$, $| \psi_i \rangle = | 0 \rangle$}
    \label{fig:blochidentity}
\end{figure}

Figure \ref{fig:blochidentity} shows the unitary path on the Bloch sphere by the control pulses generated by the EO-GRAPE algorithm (left) and the reinforcement learning agent (right). 
An initial state of $| \psi_i \rangle = | 0 \rangle$ (green vector) and a target unitary of $U_T = R_{x} (\pi/2)$, is used, resulting in a target state of $\left | \psi_T\right \rangle = \frac{1}{\sqrt{2}} (| 0 \rangle - i |1 \rangle) \equiv |-i \rangle$ (orange vector). 
The control Hamiltonian operators are $\hat{\sigma}_k \in \{ \hat{\sigma}_{x}^{1}, \hat{\sigma}_{y}^{1} \}$.
To clearly visualize the geodesic induced by the control pulses, the drift Hamiltonian is turned off (i.e., set to identity) $\hat{H}_D = I_2$. 
One can see in figure\ref{fig:blochidentity1} that the path induced by the control pulses generated by the EO-GRAPE algorithm is much more smooth and straight than the path induced by the control pulses generated by the reinforcement learning agent depicted in figure~\ref{fig:blochidentity2}. 
This is exemplary of how the reinforcement learning agent learns by reward, as seen by the random walk that the state vector travels before eventually arriving close to the target state.

\begin{figure}[!bth]
     \centering
     \begin{subfigure}[b]{0.48\textwidth}
        \centering 
        \includegraphics[clip, trim=2.4cm 2.4cm 1.7cm 1.3cm, width=0.60\textwidth]{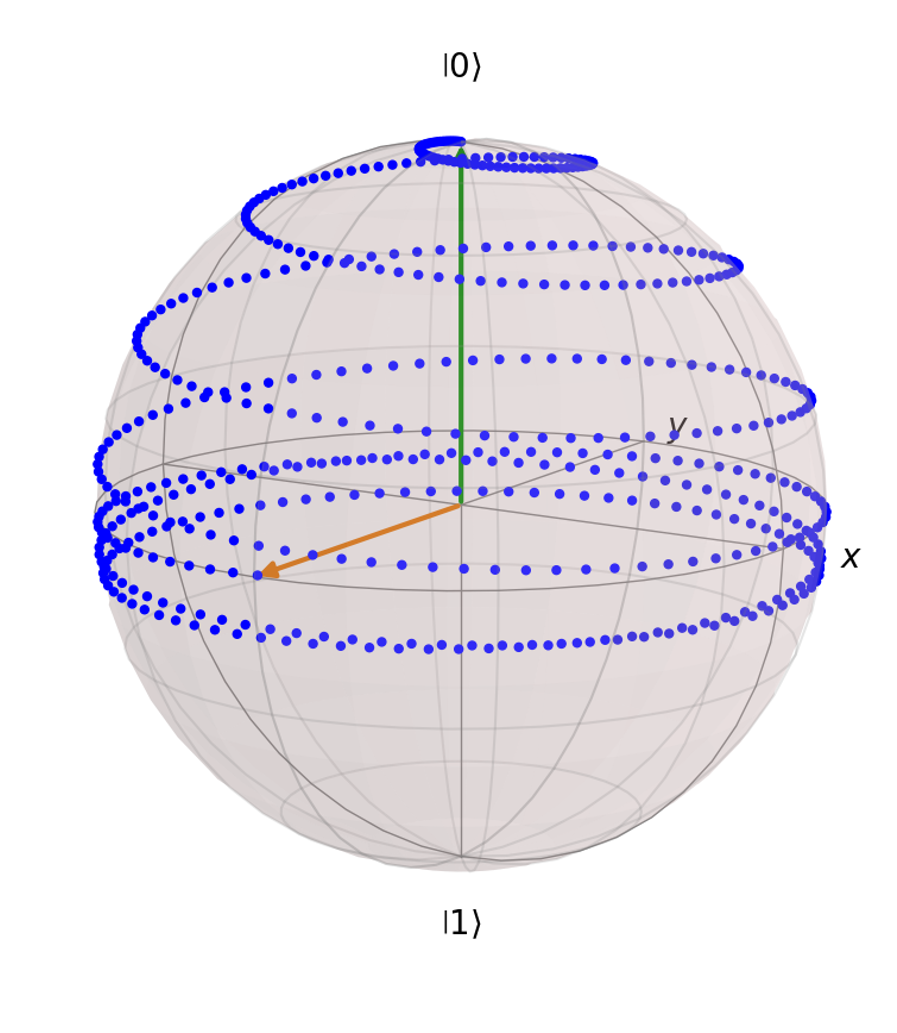}
         \caption{}
         \label{fig:blochsigmaz1}
     \end{subfigure}
     \begin{subfigure}[b]{0.48\textwidth}
        \centering 
        \includegraphics[clip, trim=1.9cm 4.8cm 3.2cm 1.9cm, width=0.60\textwidth]{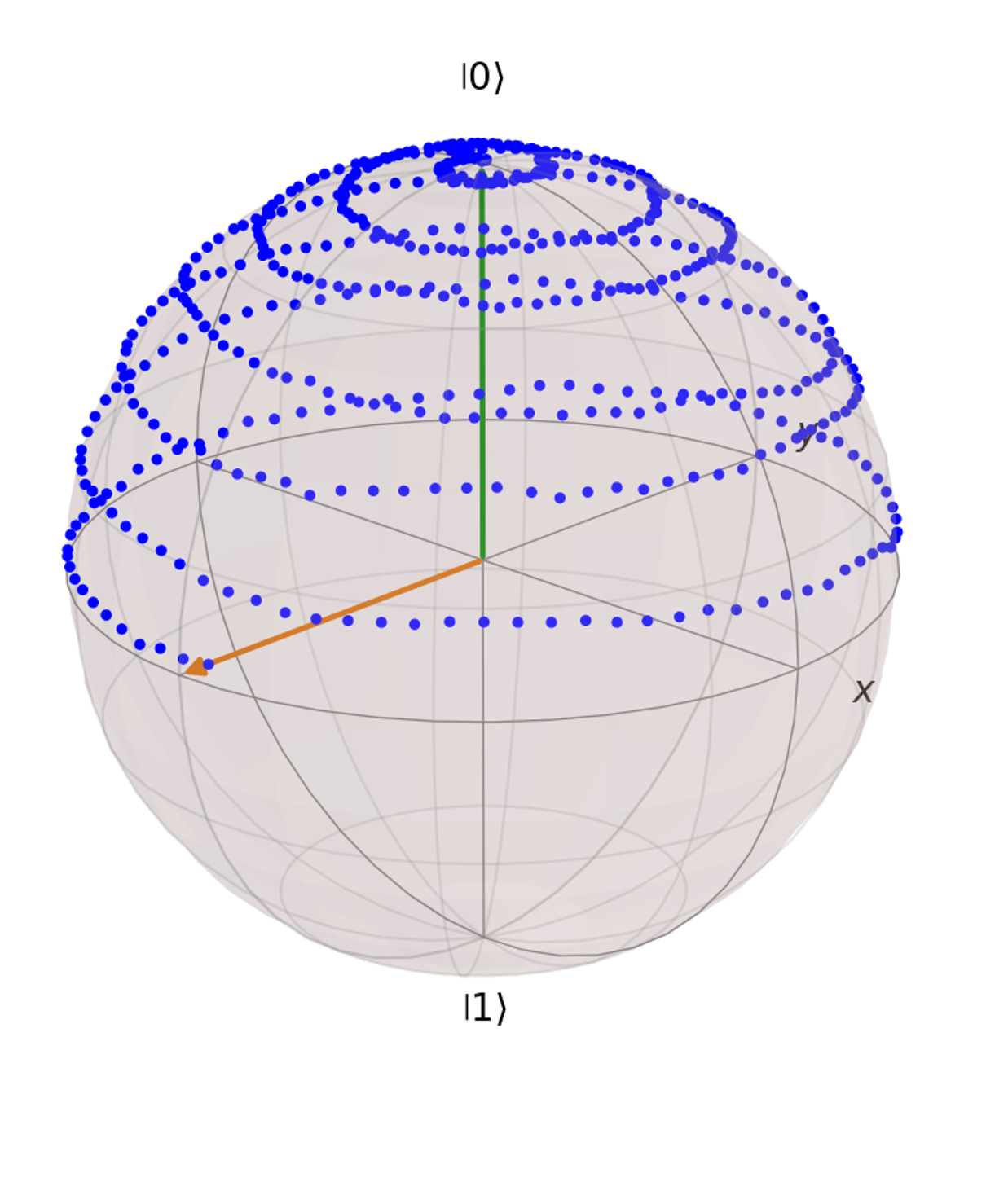}
         \caption{}
         \label{fig:blochsigmaz2}
     \end{subfigure}
    \caption{Initial state (green vector), target state (orange vector), and path of the quantum unitary (blue) in the presence of a drift Hamiltonian $\hat{H}_D$, by \textbf{(a)} EO-GRAPE generated pulses and \textbf{(b)} RL generated pulses on the Bloch Sphere. Parameters: $U_T = R_{x} (\pi/2)$, $\hat{H}_D = \hat{H}_D^{1}$, $\hat{\sigma}_k \in \{ \hat{\sigma}_{x}^{1}, \hat{\sigma}_{y}^{1} \}$, $T_1 = 1000T$, $T_2 = 1000T$, $w_f = 1$, $w_e = 0$, $N = 500$, $N_G = 500$, $N_{QRLA} = 10000$, $| \psi_i \rangle = | 0 \rangle$}
    \label{fig:blochsigmaz}
\end{figure}

In figure \ref{fig:blochsigmaz}, the unitary path on the Bloch sphere by the control pulses generated by the EO-GRAPE algorithm (left) and the reinforcement learning agent (right) are shown in the presence of a drift Hamiltonian $\hat{H}_D = \hbar \omega_1 / 2 \hat{\sigma}_z$. 
The same initial state, target quantum unitary, and control Hamiltonian operators are used. 
One can see the effect of the drift Hamiltonian, causing the state vector to precess about the $\hat{z}$-axis at the qubit frequency $\omega_1$. 
Interestingly, one can see that the path induced by the control pulses generated by the EO-GRAPE algorithm first overshoots the target state and then rotates back up to reach the target state, while the path induced by the control pulses generated by the reinforcement learning agent does not overshoot and arrive at the target state in one time. 
We suspect that this is due to the restriction imposed by the architecture of the GRAPE algorithm on the energy difference between consecutive time slices of the pulse.
This limitation can potentially be harnessed by EO-DRLPE to optimize the path length (and thus the energetic cost) further over that of EO-GRAPE.

\begin{figure}[bth]
    \centering
    \includegraphics[width = 0.55\linewidth]{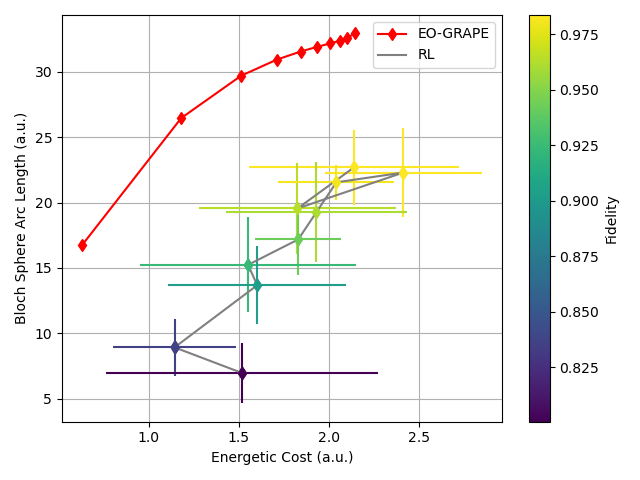}
    \caption{Combined plot of EO-GRAPE and EO-DRLPE generated pulses showing the correlation between energetic cost and path length of the unitary on the Bloch Sphere, where the color coding indicates the average fidelity of the control pulse. Each diamond marker corresponds to a different weight setting (in steps of $0.1$, with lines joining them in ascending order of $w_e$. Parameters: $U_T = R_{x} (\pi/2)$, $\hat{H}_D = \hat{H}_D^{1}$, $\hat{\sigma}_k \in \{ \hat{\sigma}_{x}^{1}, \hat{\sigma}_{y}^{1} \}$, $T_1 = 1000T$, $T_2 = 1000T$, $w_f = [1, 0.1]$, $w_e = [0.1, 0.9]$, $N = 100$, $N_{QRLA} = 10000$, $| \psi_i \rangle = | 0 \rangle$}
    \label{fig:combinedgraperlarclength}
\end{figure}

The theory presented in \cite{deffner2021energeticcost} suggests that the most energy-efficient quantum unitary is equivalent to a state vector traveling from the initial state to the target state via the geodesic between the two states on the Bloch sphere surface. 
In this scenario, the path induced by the reinforcement learning agent seems shorter than the path induced by the EO-GRAPE algorithm, suggesting that the energetic cost of the reinforcement learning agent control pulses is lower than the EO-GRAPE generated pulses in this specific scenario. 
To further investigate this, we explore the relation between the energetic cost and the path length of the unitary path on the Bloch sphere. 
In figure \ref{fig:combinedgraperlarclength}, the correlation between the path length of the path on the Bloch sphere and the energetic cost of the control pulses of both EO-GRAPE generated pulses and RL generated pulses are shown. 
The path length was calculated by summing all the small line segments around the arc of the Bloch Sphere between each time step (blue dots in figure \ref{fig:blochsigmaz}).  

One can see that for a weight setting of $w_e = 1$, $w_f = 0$, the pulses generated by the EO-GRAPE algorithm give a lower energetic cost due to the more structured nature of the optimizer. 
However, for lower weight settings of the energetic cost, the EO-DRLPE is able to find pulses with a much shorter path length than the EO-GRAPE-generated pulses.
The likely reason is that the RL agent is not restricted to keeping the pulse harmonics in shape.

As can be seen from the results presented in this subsection, the theory presented in section \ref{sec:eoc_theory} agrees with our findings. 
The shorter the path length on the Bloch sphere, the lower the energetic cost required to implement a quantum unitary gate. 
Furthermore, the control pulses generated by the EO-GRAPE algorithm take some form of accessible geodesics, as described by the theory in \cite{deffner2021energeticcost}.
In our opinion, benchmarking quantum control techniques based on the path traversed on the Bloch sphere, or more generally, the quantum geometric tensor~\cite{meinersen2024quantum}, is a promising research direction.

\section{Conclusion} \label{sec:conclusion}

In this research, we focus on estimating the energetic cost of synthesizing a quantum unitary gate. 
In this regard, this research addresses the relation between fidelity and the energetic cost and presents two novel quantum optimal control strategies to co-optimize fidelity and the energetic cost for pulse engineering.

We show that the energetic cost of implementing a quantum unitary gate through discrete pulse level control can be quantified by integrating the norm of the total Hamiltonian required to implement a specific quantum unitary gate over the total gate duration. 
In addition to this, we have found that this energetic cost positively correlates to the path length of the quantum unitary on the Bloch sphere, supporting the theory that the most energy-efficient way to implement a quantum unitary gate is through the geodesic between two quantum states. 
We demonstrate an empirical Pareto frontier between the fidelity and the energetic cost required to implement a quantum unitary gate. 
Specifically, in our experimental setup, we show that a decrease in the energetic cost by 10\% yields an increase in the infidelity of roughly 1\% in the low infidelity range. 

To develop a quantum optimal control strategy, we developed a novel cost function and gradient for the GRAPE method, allowing for co-optimization of both the fidelity and energetic cost of a quantum unitary gate. 
This novel algorithm is the energy-optimized gradient ascent pulse engineering~(EO-GRAPE). Next to the gradient-based open-loop quantum optimal control method, we have also investigated a learning-based, model-free, closed-loop method. 
An energy-optimized deep reinforcement learning for pulse engineering~(EO-DRLPE) architecture was developed to interact with a quantum environment and to learn control pulses that minimize both the energetic cost as well as the infidelity. 
Our results demonstrate that both optimal control methods perform relatively well in low-noise systems. 
When one decreases the relaxation and decoherence times of the qubits in the system, the EO-GRAPE algorithm outperforms the reinforcement learning agents for all noise settings and neural network sizes. 
Finally, a positive proportional relation between the path length of the quantum unitary on the Bloch sphere and the energetic cost of a control pulse sequence was observed, suggesting that the notion of energy efficiency and geodesics on the Bloch sphere is correct and could be leveraged to benchmark quantum control strategies.  
An open-sourced Python software package has been developed, which implements the methods discussed in this article.

More generally, this work demonstrates that one can co-optimize a quantum unitary gate on energy efficiency as well as fidelity by using quantum optimal control methods within a quantum compiler.

\subsection{Future directions}

Here we outline some promising directions for extending the research presented in this article.

In this work, we have investigated co-optimizing control pulses on both fidelity and energetic cost for gates from the universal gate set of CNOT, Hadamard, and T. 
However, one could also train a reinforcement learning agent to learn and create other energy-optimal pulses that create new universal gate sets from a specific set of hardware restrictions, such as a specific waveform, set of frequencies, or bandwidth~\cite{conceptlearning}. 
This removes the restriction of specific quantum gate synthesis and allows the agent to optimize over the set of unitaries that together form a universal gate~\cite{sarkar2024yaqq}, and thereby any quantum computation, with a potentially lower energetic cost (than using CNOT, Hadamard, T). 
 
The algorithms developed in this work generate and train on control pulses that are represented as 2D matrices with dimensions $(K, N)$. 
If one has $3$ control operators and $500$ time steps, a control pulse already contains $1500$ individual parameters that the algorithm and reinforcement learning agents need to adjust. 
However, as one can see from the pulses generated by the EO-GRAPE algorithm, the pulses can often be decomposed into individual $\sin$ or $\cos$ periodic functions. 
It is, therefore, plausible that one can transform the control pulses to the frequency domain by applying a Fourier transformation, or using alternate learning basis like Kolmogorov-Arnold networks~\cite{sarkar2024kanqoc}. 
This would allow for representing a control pulse with just a few amplitude and frequency parameters~\cite{frequencyoptimization}. 
This would dramatically increase the computational efficiency of both algorithms and could also increase the performance of the reinforcement learning agent, as it would automatically apply harmonic pulses instead of random block pulses. 

Different formulations of the drift and control Hamiltonian operators have been investigated in this research. 
However, the effect of having multiple redundant control operators has not been thoroughly researched. 
Minimizing the number of control operators per qubit to universally control the qubits is, therefore, a critical component in minimizing the energetic cost of a quantum unitary gate. 
We can, therefore, ask ourselves, given a multi-qubit system with drift Hamiltonian $\hat{H}_D$, how many individual control terms $|\hat{H}_k|$ does one need to control the quantum system universally?
There exists a theoretical framework to address this problem of full controllability called the Lie rank test~\cite{lierank}. 
One could, therefore, use this theoretical framework to minimize the number of control operators needed per qubit and per drift Hamiltonian and observe the effect that it has on the energetic cost of quantum unitary gates performed on the qubits. 

Information, entropy, and energy are closely related quantities. 
Landauer's principle states that the minimum energy required by a logic operation will be the temperature times the entropy \cite{landauer}. 
As one can see, the entropy (expressed as bits of information) and energy of a control sequence are closely related~\cite{shannon}. 
Therefore, a similar investigation as the one performed in this research between fidelity and energy can be proposed for information and energy. 
One can investigate the co-optimization of both the information contained in a control sequence~\cite{mishra2024}, as well as the energy of a control sequence, and see what the relation is between the two and if it is possible to co-optimize, similar to fidelity and energy. 
In theory, one could also change the fidelity part of the cost function presented in this work by measuring for information and repeating the same experiments that have been performed in this work.

\section*{Software availability} \label{software}

The open-sourced code for the project, configuration files, output data, and plotting codes for the experiments presented in this article are available at:
\href{https://github.com/QML-Group/EO-QCtrl}{https://github.com/QML-Group/EO-QCtrl}.

\section*{Acknowledgements}

A didactic introduction and additional details of the implementation presented in this article can be found in the corresponding master thesis~\cite{SebastiaanMSc}.
We thank Ramon Overwater, İlker Polat, and Yining Zhu for insightful discussions on the current state-of-the-art pulse control optimization for superconducting qubit processors in QuTech.
AS acknowledges funding from the Dutch Research Council (NWO) through the project ``QuTech Part III Application-based research" (project no. 601.QT.001 Part III-C—NISQ).

\bibliographystyle{unsrt}
\bibliography{ref.bib}


\end{document}